\documentclass[letterpaper,twocolumn,10pt]{article}
\usepackage{usenix,epsfig,endnotes}
\usepackage{todonotes}
\usepackage{enumitem}
\usepackage{subcaption}
\usepackage{array}
\usepackage{multirow}
\usepackage{graphicx}

\usepackage{tikz, pgfplots}
\usetikzlibrary{trees, arrows}
\usepgfplotslibrary{external}
\begin{document}

\newcommand{\AM}[1]{\textcolor{purple}{AM: #1}}
\newcommand{\GG}[1]{\textcolor{blue}{GG: #1}}
\newcommand{\gf}[1]{\textcolor{orange}{GF: #1}}
\newcommand{\lc}[1]{\textcolor{red}{LC: #1}}
\newcommand{\newrevision}[1]{{#1}}
\newcommand{\newnewrevision}[1]{\textcolor{red}{#1}}

\renewcommand{\paragraph}[1]{\vspace{0.05in}\noindent \textbf{{#1}}}

%don't want date printed
\date{}

%make title bold and 14 pt font (Latex default is non-bold, 16 pt)
\title{\Large \bf Privacy Requirements and Realities of Digital Public Goods}

\def\plainauthor{}

\newif \ifpublicversion
\publicversiontrue
% \publicversionfalse

\newif \ifarxiv
\arxivtrue
% \arxivfalse

\ifpublicversion
    \author{
    {\rm Geetika Gopi}\\
    Carnegie Mellon University
    \and
    {\rm Aadyaa Maddi}\\
    Carnegie Mellon University
    \and
    {\rm Omkhar Arasaratnam}\\
    OpenSSF
    \and
    {\rm Giulia Fanti}\\
    Carnegie Mellon University
    }
    
    \newcommand{\update}[1]{#1}
\else
    % anon for submission
    \author{}

    \newcommand{\update}[1]{\textcolor{green!50!black}{#1}}
\fi

\ifarxiv
\else
    \pagenumbering{gobble}
\fi

\maketitle
\thecopyright

% Use the following at camera-ready time to suppress page numbers.
% Comment it out when you first submit the paper for review.
\thispagestyle{empty}

\begin{abstract}
In the international development community, the term ``digital public goods'' is used to describe open-source digital products (e.g., software, datasets) that aim to address the United Nations (UN) Sustainable Development Goals. DPGs are increasingly being used to deliver government services around the world (e.g., ID management, healthcare registration). Because DPGs may handle sensitive data, the UN has established user privacy as a first-order requirement for DPGs. The privacy risks of DPGs are currently managed in part by the DPG standard, which includes a prerequisite questionnaire with questions designed to evaluate a DPG's privacy posture. 

This study examines the effectiveness of the current DPG standard for ensuring adequate privacy protections. We present a systematic assessment of responses from DPGs regarding their protections of users' privacy. We also present in-depth case studies from three widely-used DPGs to identify privacy threats and compare this to their responses to the DPG standard. Our findings reveal serious limitations in the current DPG standard's evaluation approach. We conclude by presenting preliminary recommendations and suggestions for strengthening the DPG standard as it relates to privacy. \update{Additionally, we hope this study encourages more usable privacy research on communicating privacy, not only to end users but also third-party adopters of user-facing technologies.}
\end{abstract}

\section{Introduction}
Today, digital government services---like national registries, payment systems, or healthcare systems---are often implemented and administered by third-party vendors \cite{3rdpartyservices2018}. 
This comes with a few drawbacks. 
Vendors are known to charge high prices \cite{forbes2021trends}, and governments are subsequently subject to vendor lock-in, due either to monopolies or a lack of interoperability between market offerings \cite{onezero2024tech}. 
These costs are typically passed on to residents in the form of taxation, which can be particularly problematic in low-income countries \cite{worldbank2024lowincome}.

Digital Public Goods (DPGs) are a concept that was recently revived in the international development community, partially to counter this trend. 
DPGs are open-source digital goods that are designed to benefit society \cite{digitalpublicgoods}. 
In 2020, the United Nations (UN) put forth a report  calling for ``a platform for sharing digital public goods... in a
manner that respects privacy, in areas related to attaining the Sustainable Development Goals.'' \cite{un-mandate}. 
In response, the Digital Public Goods Alliance (DPGA) was formed to encourage and steward the development of DPGs. 
Precisely, the DPGA defines DPGs as ``open source software, open data, open AI models, open standards and open content that adhere to privacy and other applicable laws and best practices, do no harm, and help attain the [United Nations' Sustainable Development Goals]'' \cite{dpga}.

In the years since the DPGA was formed, DPGs have occupied a growing role in government services worldwide, as well as other community-driven services. For example, MOSIP is a DPG digital ID system with over a 100 million users, currently adopted by 11 countries \cite{mosip}. DIGIT HCM is a health campaign management DPG with over 15 million users \cite{divoc2024homepage}. DIVOC is another health campaign management DPG with over 160 million users and is currently adopted by 5 countries including India, Philippines and Sri Lanka \cite{divoc2024homepage}. 

Privacy is a first-order concern in DPGs. In addition to being highlighted as a key property in the original UN report \cite{un-mandate}, it is also a central component of the evaluation used to select which projects are officially listed as DPGs \cite{dpga2024standard}.
Briefly, the evaluation proceeds as follows (detailed description in Section \ref{sec:background}): a DPG candidate project first submits answers to an official DPG questionnaire, which contains 24 questions \cite{dpga2024standard}.
One of these questions specifically asks about what personally identifiable information (PII) the DPG candidate collects, and how this information is protected. 
Responses from the candidate are meant to be backed up by supporting documentation and/or code. 
DPG candidates' responses to the questionnaire are  then reviewed by the DPGA; if the responses are judged to be sufficiently high-quality and consistent with the DPGA's requirements, the candidate is formally approved as a DPG. Although DPG certification does not convey any explicit rights or privileges to the holders, it appears to be used in practice as a form of advertisement, e.g., by being listed on the DPG's website \cite{mosip,aamdigital2024homepage,divoc2024homepage}. 

In this paper, we study whether the DPG approval process is effective at selecting DPGs that protect \textbf{user privacy}. We evaluate this in two phases: first, we run a qualitative study to analyze the responses of DPGs to the privacy component of the questionnaire. We evaluate these responses in terms of their completeness and their adherence to established privacy best practices. We then conduct an in-depth case study on three DPGs, in which we analyze their structure, documentation, and possible privacy threats. We use this analysis to determine whether DPGs may have privacy implications that are not captured by their responses to the DPG standard.

Our results show that existing DPGs provide a wide range of responses to the questionnaire, many of which convey limited maturity or attention  to privacy and data protection.
This suggests that the current DPG standard, and the associated approval process, does not  filter out projects that take a lax approach to privacy. 
We emphasize that the DPGA is in a difficult position: it is neither a standards agency nor an enforcement agency. It is unrealistic to expect that it will be able to evaluate the privacy properties of candidate DPGs, many of which comprise complex code bases and documentation. 
Nonetheless, the UN has outlined privacy as a first-order requirement for DPGs. 
Hence, we believe it is important to find a solution that both encourages privacy best practices from DPGs, while also working within the existing constraints.

\paragraph{Contributions}
Our contributions in this paper are threefold: 
\begin{enumerate}[leftmargin=*]
    \item We conduct a qualitative study of the privacy responses of 101 DPGs. 
    We code their responses according to (a) high-level qualitative properties, and (b)   common privacy themes that were extracted from existing privacy frameworks. We find that a high percentage (40\%  
    of DPGs) did not provide an adequate level of detail to understand how they handle PII. Moreover, many DPGs make common mistakes, such as conflating privacy compliance with privacy protection (50\%) and shifting responsibility for data stewardship to other parties (17\%). 
    \item We conduct three in-depth case studies of DPGs 
    with over 1 million users and from different sectors. Among these, we find that even mature DPGs that answered the DPG standard thoroughly can have gaps in how their documentation addresses privacy concerns. These gaps have been communicated to the relevant DPGs.
    \item We make several recommendations for how to improve the DPG standard to encourage better privacy protections. At a high level, our main recommendation consists of requiring a more detailed privacy assessment (akin to Privacy Impact Assessments \cite{PIA}), to be completed by third parties or the DPGs themselves. The DPGA would no longer evaluate the quality of privacy responses, but would provide the privacy assessment documentation on their website for downstream users to evaluate. These recommendations (outlined in more detail in Section \ref{sec:recs}) have been communicated with the DPGA, and are currently under consideration for a restructuring of the DPG standard. 
\end{enumerate}

\update{Usable privacy research often focuses on privacy for end-users. However, for DPGs, there are several stakeholders that want to (a) demonstrate that their methods are private (DPG candidates) and (b) evaluate the privacy claims of other organizations (DPGA) who also need those processes to be ``usable''. While existing research has studied how to communicate privacy to end users (e.g., through privacy nutrition labels \cite{kelley2009nutrition, li2022understanding}), little research has been conducted on communicating privacy to third-party adopters, who have more technical sophistication than a typical user, but less sophistication than a domain expert. This area is relatively under-explored, and the DPG environment, being open-source, is an excellent opportunity to study such questions.}

\section{Background and Related Work}
\label{sec:background}
\subsection{The DPG Standard and Questionnaire for Privacy}
Assessing and endorsing a DPG candidate involves a three-step process. The organization seeking DPG status must first complete an online application on the DPGA's website, including a DPG questionnaire \cite{dpgstandard-overview}. As part of the application, candidates are required to submit various forms of supporting evidence such as technical documentation, open licenses, and privacy policies. Once the application is received, it undergoes an evaluation process based on the DPG standard \cite{dpga2024standard}, which serves as a set of specifications and guidelines that defines a Digital Public Good. To receive recognition from the DPGA and the wider community, a DPG candidate must meet the baseline requirements as outlined in the DPG standard.
If the application satisfies all the criteria of the DPG standard, it will be acknowledged as a Digital Public Good and included in the DPG registry \cite{dgpa2024registry}.

The evaluation process for the DPG standard involves a thorough assessment of various criteria including accessibility, functionality, interoperability, and privacy, among others. The DPG standard is an open-source standard maintained by the DPGA \cite{dpga2024standard}. Its credibility is further 
endorsed by a growing list of experts who advocate for open-source entities \cite{dpga2024standard}.

The DPG standard \cite{dpga2024standard} includes three privacy-related sections, i.e., Sections 7, 8 and 9, outlining the privacy requirements for a DPG candidate. Section 9(a) specifically addresses data privacy and security by requiring DPGs to demonstrate how they ensure the privacy, security and integrity of personal information collected, stored and distributed as part of their solution. Section 7 asks candidates to explain how they ensure compliance with relevant privacy and applicable laws. Sections 8 and 9 of the DPG standard further require candidates to explain their efforts to follow best practices and ensure that the solution does no harm to their users.

The DPG standard is implemented via the DPG questionnaire \cite{dpga2024standard}, which evaluates candidates' adherence to the 9 indicators of the DPG standard. The questionnaire comprises both open-ended and multiple-choice questions. In this study, we are most interested in Section 9(a), which requires candidates to respond to the following question:

\begin{quote}
    ``How does your solution ensure data privacy \& security? Please demonstrate how the project ensures the privacy, security and integrity of this data and the steps taken to prevent adverse impacts resulting from its collection, storage and distribution.'' (open form)
\end{quote}

This open-ended question leaves much room for interpretation, and does not precisely define what is meant by privacy. Hence, our primary research question for this paper is as follows:\\

\textit{Does the current DPG standard effectively evaluate or document digital solutions'  potential privacy harms?}

\bigskip
To study this question further, we first conducted a qualitative analysis of 101 DPG responses to the privacy component of the questionnaire. This is described further in Section \ref{sec:qual_analysis}. 
We then conducted an in-depth case study on three DPGs to elicit possible privacy threats that the DPG questionnaire fails to capture. This is described further in Section \ref{sec:case_studies}.

\subsection{Related Work}
\label{sec:related}

\paragraph{Evaluating the Potential and Drawbacks of DPGs}
While DPGs have not received as much attention from the academic research community, several papers (many of them position papers)  highlight the significance of DPGs and the factors that impact their utility \cite{nicholson2022agenda,nicholson2022digital,kummer2020unemployment,saebo2021digital,eaves2022best,gillwald2019governance,tavares2023digital,sturmer2023digital,mukherjee2021fast,chen2023motivating}. Nickholson \emph{et al.} explored key challenges and opportunities in achieving the Sustainable Development Goals through DPGs \cite{nicholson2022agenda}. Their writing emphasizes that the potential harm from a DPG is not only associated with the technology itself, but also depends on its implementation, usage, and evolution over time, highlighting the need for further research in the DPG space \cite{nicholson2022agenda}. 
This observation closely aligns with our own findings, and partially motivates this study.
Mukherjee \emph{et al.} describe case studies that illustrate the concept of digital building blocks as public goods and demonstrate their application to developmental challenges such as poverty, inequality, health, education, public administration, and governance that affect entire populations \cite{mukherjee2021fast}. Kumar et al. and Chen et al. explore the factors influencing contributions to DPGs, while also highlighting the importance of enhancing the quality of DPGs \cite{kummer2020unemployment,chen2023motivating}.
These studies conducted large-scale field experiments and employed power analysis methods to study the correlation between factors influencing experts' contributions to DPGs.

\paragraph{Incentivizing Privacy Best Practices}
While there has been a vast literature studying how to incentivize organizations to invest in cybersecurity \cite{lelarge2009economic,kelly2012investing,rosson2019incentivizing,fedele2022dangerous}, there has been comparatively less work analyzing economic incentives of organizations to invest in data privacy 

\cite{adjerid2016impact,lam2019data,wright2016enforcing}. 
Instead, many organizations' policies and procedures surrounding data privacy are primarily driven by compliance with privacy regulation, either directly \cite{voigt2017eu,act1996health,gramm1999gramm,coppa,wright2016enforcing,sirur2018we} or indirectly, e.g., via vendor requirements \cite{PIA,ashley2002privacy}.
However, compliance with privacy regulations \emph{does not inherently ensure that an organization is adequately protecting user privacy} \cite{gray2007privacy}. 
While there is not a single global standard for data privacy, many existing privacy frameworks (e.g., LINDDUN \cite{wuyts2015linddun}) present compliance as only one part of a robust privacy posture \cite{nist2024privacyframework,PIA}. 
Indeed, empirical observations show how various components of a privacy strategy, beyond just compliance, can interact to affect the utility of a product. 

For instance, Adjerid \emph{et al.} showed that privacy regulation combined with collecting proper consent from users can actually result in greater data sharing than under fully unregulated situations \cite{adjerid2016impact}. This suggests the importance of coupling structured privacy requirements around privacy with a clear mechanism for collecting user consent. 
At the same time, the very act of asking for consent can affect users' willingess to share their data with a service, as demonstrated by Lam \emph{et al.} with regards to the opt-in requirement of GDPR \cite{lam2019data}.

While our recommendations in Section \ref{sec:recs} relate to incentivizing privacy best practices, this paper focuses on the higher-level question of whether the current DPGA standard ensures that  DPGs' privacy postures are consistent with the recommendations of prominent privacy frameworks.

\section{Methodology}
Our evaluation is split into two components. 
\begin{enumerate}[leftmargin=*]
    \item \textbf{Qualitative Analysis of DPG Responses (\S\ref{sec:qual_analysis})} We first conducted a qualitative document analysis of DPG responses from all approved DPGs as of May 12, 2023.
    Our goal was to understand trends in the content and quality of DPG candidates' responses to the privacy question. 
    \item \textbf{DPG Case Studies (\S\ref{sec:case_studies})} We next conducted detailed case studies into three DPGs to understand their privacy implications. The case studies were conducted on August, November, and December 2023, respectively.
    Our goal for this component was to understand how responses in the DPG standard are correlated (or not) with actual implementations or architectures.
\end{enumerate}
We next detail the methodology for each component. We discuss the limitations of our methodology in Section \ref{sec:limitations}.

\subsection{Methods: Qualitative Analysis of DPG Responses}
\label{sec:methods-qual-analysis}
To analyze DPG responses, we first gathered all 167 DPG responses from the DPGA's GitHub repository on May 12, 2023 \cite{dpgalliance2024repository} and filtered them to include only DPGs indicating the collection of personally identifiable information (PII)\update{, as these are the only ones that answer Section 9(a)}. This resulted in a total of 101 relevant DPG responses. 
Filtering was needed because DPG candidates that did not indicate collection of PII would have no further statements to analyze regarding privacy. \update{We could not access rejected DPG responses.}

\begin{figure*}[t]
    \centering
    \scalebox{.75}{\begin{tikzpicture}[sibling distance=25mm]
  \node[text width=20mm] {Proposed Protection Mechanisms}[sibling distance=25mm]
    child {node[text width=15mm] {Data\\ Oriented}
        child {node[text width=25mm, xshift=-12mm, yshift=-5mm] {Commercial/In-house Tools, Data Strategies, Secure Data Storage}}
    }
    child {node[text width=15mm] {Process Oriented}
      child {node[text width=20mm, yshift=-13mm] {Access Control, Security Upgrades/Patches, Vulnerability Testing, Strong Passwords}}
      child {node[text width=20mm, yshift=-10mm] {Notify 3rd Party Data Sharing, Governance Processes/Audits, User Control}}
    };

    \begin{scope}[xshift=6cm]
    \node[text width=25mm] {Provided Supporting Material}[sibling distance=15mm]
    child {node[text width=15mm] {End-User Focused}
        child {node[text width=15mm, yshift=-5mm] {Cookie Policy, Privacy Policy}}
    }
    child {node[text width=15mm, xshift=5mm] {Adopter Focused}[sibling distance=15mm]
      child {node[text width=15mm, yshift=-2mm] {Security/\\Privacy Docs}}
      child {node[text width=15mm] {Compliance Docs}}
    }
    child {node[xshift=3mm, yshift=2mm] {No Docs}};
    \end{scope}

    \begin{scope}[xshift=11cm]
    \node[text width=20mm] {Overall\\ Response\\ Quality}
    child {node[text width=15mm, yshift=-2.5mm] {Addresses Privacy/\\Security}
        child {node[text width=25mm, yshift=-10mm, xshift=-8mm] {Security-related Privacy, Security Only, Partially Addresses Privacy}}
    }
    child {node[text width=15mm] {Unclear\\ Response}[sibling distance=15mm]
        child {node[text width=20mm, yshift=-15mm, xshift=-2mm] {Lack of Specificity, Clarifies Data Ownership, Downplaying Risks, Compliance Implies Protection}}
        child {node[text width=20mm, yshift=-8mm, xshift=7mm] {Unclear PII Collection, Inconsistent Answer, Does Not Answer Question}}
    };
    \end{scope}

    \begin{scope}[xshift=16cm]
    \node {Privacy Component Analysis}%[sibling distance=25mm]
    %child {node[text width=25mm, xshift=15mm, yshift=-17mm] {Security Safeguards, Regulatory Efforts, Privacy by Design, Notice and Consent}}
    child {node[text width=25mm, xshift=10mm, yshift=-15mm]  {User Choice, Data Use Limitation, Data Collection Limitation, Data Accuracy, Security Safeguards, Regulatory Efforts, Privacy by Design, Notice and Consent}};
    \end{scope}
\end{tikzpicture}}
    \caption{Categorizing codes under the four themes we consider during qualitative analysis of DPG responses.}
    \label{fig:enter-label}
\end{figure*}
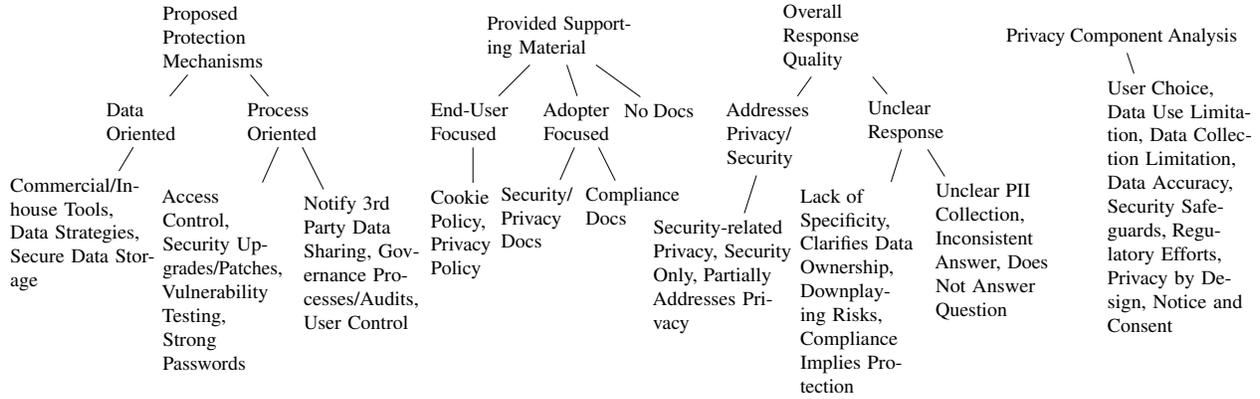

\paragraph{Analysis}
\update{The lead researcher coded 50 responses independently to develop the initial codebook. The lead researcher developed ‘Privacy Component Analysis’ codes with a priori coding, using existing privacy frameworks (e.g., LINDDUN, APEC) \cite{wuyts2015linddun, cooperation2005apec}. For remaining themes, the lead researcher used emergent coding. The lead and second researcher went over the coded responses and refined the codebook through discussions. The two coders coded the remaining 51 responses independently using the final codebook (Appendix A). The inter-rater reliability (IRR) was computed over these responses using percentage agreement (responses coded the same way, divided by the total number of responses). We achieved an inter-rater reliability of 0.87, which is considered acceptable \cite{o2020intercoder}. Since codes were not mutually exclusive, Cohen’s kappa was inapplicable. Conflicts were resolved through discussion.}
We emphasize that our study size is relatively small (101 DPGs), but consists of the \emph{entire} population of DPGs that claimed to collect PII at the time of data collection. Hence, we present counts of occurrences of codes.

\subsection{Methods: Detailed Case Studies}
For our case studies, we chose three DPGs based on specific criteria: (1) having a user base of over 1 million users, and (2) providing documentation with specific sections related to privacy. 
Since these large-scale systems are labor-intensive to review and analyze, we decided to focus on 3 DPGs with significant impact in user-facing sectors: healthcare, digital IDs, and news and media. 

\paragraph{Analysis}
Our case studies used privacy threat modeling \cite{wuyts2015linddun}, a structured approach used to identify potential privacy threats within a system or application. This involves analyzing the system's components, data flows, and potential vulnerabilities that could compromise user privacy. 
Using these techniques, we compared our findings with the responses provided by DPGs to the DPG standard. Our goal is to understand how well responses to the questionnaire relate to a more detailed analysis of the DPG. 
Below, we outline our three-step methodology for identifying privacy threats.

We first reviewed the technical documentation from the selected DPGs and identified all system components involved in processing personal data. This step gave us a thorough understanding of the DPG's system architecture. Next, we carefully identified and analyzed the data flows \cite{microsoft2006uncover} by creating Level 2 data flow diagrams \cite{level2dfds}. 
Using Level 2 diagrams lets us capture potential privacy risks without considering low-level system details. Finally, we used the LINDDUN threat modeling framework \cite{wuyts2015linddun} to identify potential privacy threats.
We highlight that this methodology is capturing potential privacy vulnerabilities that are implied by the documentation. It does \emph{not} necessarily imply that a vulnerability actually exists in the software. 
For example, some of the vulnerabilities we found were confirmed to be documentation mistakes (not true vulnerabilities) by the DPGs. 

The case study results provided valuable insights into the actual privacy practices and strategies employed by the selected DPGs, shedding light on the effectiveness of their privacy protection mechanisms. We compared this analysis to DPGs' responses on the questionnaire to explore whether DPGs can have privacy implications that
are not captured by the DPG standard.

\section{Qualitative Analysis of DPG Responses}
\label{sec:qual_analysis}

When analyzing the responses of DPG candidates to Section 9(a) of the DPG questionnaire, we generated codes related to four main themes \update{(codebook construction in \S\ref{sec:methods-qual-analysis})}:
\begin{itemize}[itemsep=0.5pt, topsep=0pt,leftmargin=*]
    \item \textbf{Overall response quality. } Were the responses clear, internally consistent, and specific? 
    \item \textbf{Types of supporting documentation.} What kind of supporting documentation did the DPG candidate provide?
    \item \textbf{Proposed privacy safeguards. } What technical and process strategies were used by DPGs to protect user data? 
    \item \textbf{Coverage of privacy best practices.} Did the response cover common elements of existing privacy frameworks and principles? 
\end{itemize}
\update{These themes were identified using a top-down approach to answer two questions: (1) How did the candidates respond — both in their main response (``Proposed privacy safeguards'') and ``Types of supporting documentation'', and 2) How well did they respond, in form (``Overall response quality'') and function (``Coverage of privacy best practices'').}

These themes helped us understand how DPGs approach privacy and whether the current evaluation process helps the DPGA screen out DPGs with possible privacy threats. A categorization of our identified codes within the four themes is illustrated in Figure \ref{fig:enter-label}. We next present our results, divided according to theme.

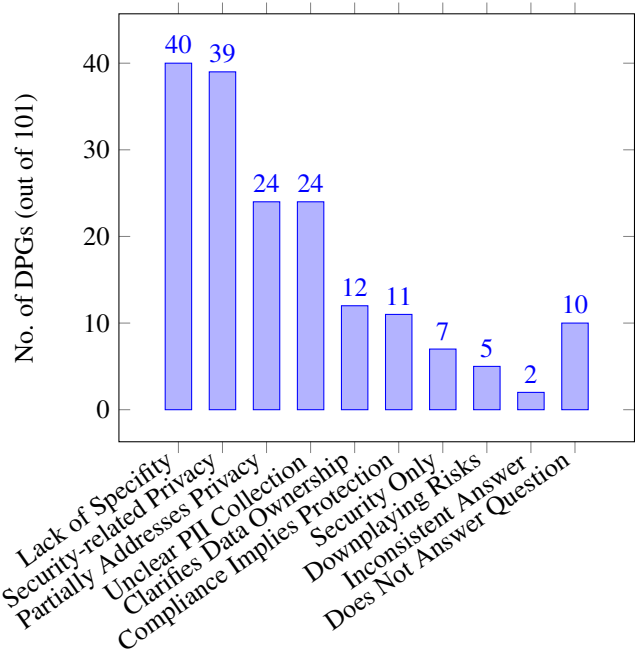
\begin{figure}[t]
    \centering
    \begin{tikzpicture}
\begin{axis}[
ybar,
enlargelimits=0.15,
legend style={at={(0.5,-0.2)},
anchor=north,legend columns=-1},
ylabel={No. of DPGs (out of 101)},
symbolic x coords={Lack of Specifity,
Security-related Privacy,
Partially Addresses Privacy,
Unclear PII Collection,
Clarifies Data Ownership,
Compliance Implies Protection,
Security Only,
Downplaying Risks,
Inconsistent Answer,
Does Not Answer Question,
},
xtick=data,
nodes near coords,
nodes near coords align={vertical},
x tick label style={rotate=35,anchor=east},
]
\addplot coordinates {
(Lack of Specifity, 40)
(Security-related Privacy, 39)
(Partially Addresses Privacy, 24)
(Unclear PII Collection, 24)
(Clarifies Data Ownership, 12)
(Compliance Implies Protection, 11)
(Security Only, 7)
(Downplaying Risks, 5)
(Inconsistent Answer, 2)
(Does Not Answer Question, 10)
};
\end{axis}
\end{tikzpicture}
    \caption{Results from qualitative analysis of DPG responses for Overall Response Quality.}
    \label{fig:org}
    \vspace{-0.2in}
\end{figure}
\subsection{Overall Response Quality}
When evaluating the overall quality of responses, we observed that the majority were either vague, solely focused on security controls, or only partially addressed privacy controls. Roughly, we categorized the responses as ones that `address security/privacy' or were `unclear responses'. Our results for overall response quality are illustrated in Figure \ref{fig:org}. 

Notably, we found that 40\% of DPGs lacked specificity in describing their protection methods. We defined `lack of specificity' as responses that mentioned privacy-related terms such as `anonymization' or `obfuscation', without explaining how it is applied within the context of the DPG solution. This could be attributed to the open-ended nature of the questionnaire. Refer to the codebook in Table \ref{tab:org_codebook} for the full list of response types and their definitions.
An example of a response coded as lacking specificity is as follows:

\begin{quote}
    \textbf{DPG15}: ``The solution promotes best security and quality assurance practices in an effort to support the privacy of PII and prevent adverse impact related to PII. Security and quality assurance best practices that can contribute to the prevention of adverse impact related to PII are integrated into our development processes and automated as possible.''
\end{quote}

Furthermore, 50\% of DPGs appeared to primarily emphasize security-related privacy controls such as encryption, \newrevision{hashing}, or regulatory measures as their main privacy strategies:
\begin{quote}
    \textbf{DPG43}: ``All data transfer through HTTPS (SSL) \& user level security is maintained through SHA-512 encryption with roles \& privileges.''
\end{quote}

\begin{quote}
    \textbf{DPG80}: ``While we don't collect the data ourselves, the software has high degrees of security and compliance at the software and network level to ensure data integrity.''
\end{quote}
While security measures and best practices are useful, they are not sufficient for guaranteeing data privacy. 
Interestingly, there was little mention of data-oriented strategies like data minimization and anonymization.

Among the stronger responses, about 25\% of responses took steps that fully or partially addressed privacy by design principles. 
For instance:
 \begin{quote}
     \textbf{DPG81}: ``We are collecting anonymized data (clinical data of patients) with prior approvals and clearances from hospitals.''
 \end{quote}
 
 On the other hand, we encountered several responses that showed a lack of understanding of privacy by design. 17\% of DPGs appear  to shift the responsibility for privacy to solution implementers;  others downplayed privacy  risk:
 \begin{quote}
     \textbf{DPG101}: ``... Unfortunately, there is no such thing as true data protection, even when data is locally stored and hosted in a country...''
 \end{quote}

 \begin{quote}
     \textbf{DPG11}: ``As a default, this project does not collect or store PII data, but some partners and deployments would like the option to have the same; in which case we store the name, address and phone number of the consenting individual and clearly mention in our contracting terms that we do not own any of this data''
 \end{quote}
Overall, these responses reveal a wide spectrum of  qualities in responses. Most importantly, they suggest that \textbf{many DPGs are currently not providing enough detail for the DPGA or an adopter to understand its privacy posture. }

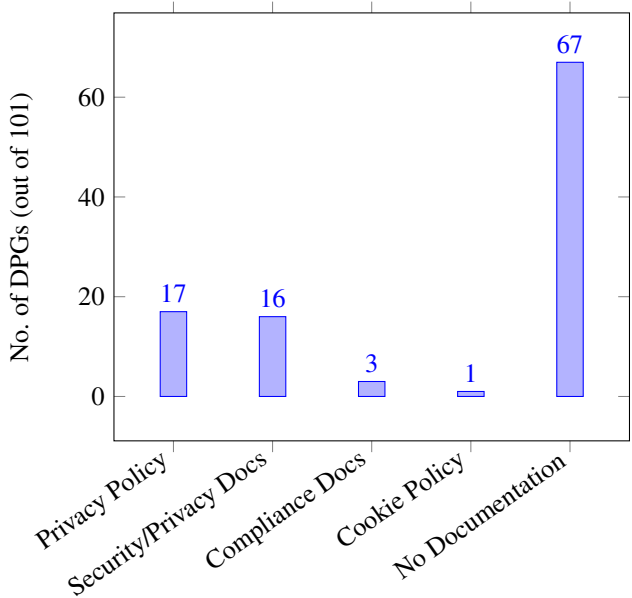
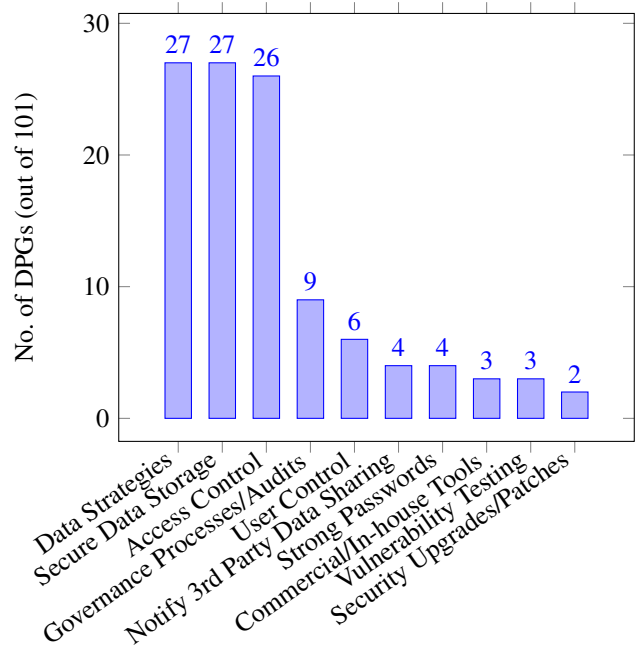
\begin{figure*}[t]
    \begin{subfigure}[b]{0.5\textwidth}
    \begin{tikzpicture}
\begin{axis}[
ybar,
enlargelimits=0.15,
legend style={at={(0.5,-0.2)},
anchor=north,legend columns=-1},
ylabel={No. of DPGs (out of 101)},
symbolic x coords={Privacy Policy, Security/Privacy Docs, Compliance Docs, Cookie Policy, No Documentation},
xtick=data,
nodes near coords,
nodes near coords align={vertical},
x tick label style={rotate=35,anchor=east},
]
\addplot coordinates {
(Privacy Policy, 17)
(Security/Privacy Docs, 16)
(Compliance Docs, 3)
(Cookie Policy, 1)
(No Documentation, 67)
};
\end{axis}
\end{tikzpicture}
    \caption{Provided Supporting Material}
    \label{fig:psg}    
    \end{subfigure}
    ~
    \begin{subfigure}[b]{0.5\textwidth}
        \begin{tikzpicture}
\begin{axis}[
ybar,
enlargelimits=0.15,
legend style={at={(0.5,-0.2)},
anchor=north,legend columns=-1},
ylabel={No. of DPGs (out of 101)},
symbolic x coords={Data Strategies, Secure Data Storage, Access Control, Governance Processes/Audits, User Control, Notify 3rd Party Data Sharing, Strong Passwords, Commercial/In-house Tools, Vulnerability Testing, Security Upgrades/Patches},
xtick=data,
nodes near coords,
nodes near coords align={vertical},
x tick label style={rotate=35,anchor=east},
]
\addplot coordinates {
(Data Strategies, 27)
(Secure Data Storage, 27)
(Access Control, 26)
(Governance Processes/Audits, 9)
(User Control, 6)
(Notify 3rd Party Data Sharing, 4)
(Strong Passwords, 4)
(Commercial/In-house Tools, 3)
(Vulnerability Testing, 3)
(Security Upgrades/Patches, 2)
};
\end{axis}
\end{tikzpicture}
    \caption{Proposed Protection Mechanisms} 
    \label{fig:ppg}
    \end{subfigure}
    \caption{Results from qualitative analysis of DPG responses.}
    \vspace{-0.2in}
\end{figure*}

\subsection{Provided Supporting Material}
\label{sec:supporting-material}
We found that few DPGs provided supporting documentation, and those that did often provided policies. We categorize these as `end-user focused' or `adopter focused' material. End-user focused documentation refers to material that is seen by individuals whose PII can be handled by the DPG. Adopter focused documentation, on the other hand, can be technical material informing DPG implementers how to use the supported security and privacy measures, or contain instructions for compliance with privacy regulations such as GDPR. 

As shown in Figure \ref{fig:psg}, over 50\% of the studied DPGs did not provide any supporting documentation to explain their protection mechanisms, 
% Despite this, they still managed to pass the DPG standard check. 
and only about 16\% of  DPGs submitted some form of documentation related to security or privacy. 
The remaining DPGs submitted privacy policies, cookie policies, or compliance-related documentation. These results suggest that some DPGs may equate privacy compliance with privacy protection. 
They focused more on demonstrating compliance with regulations, rather than implementing robust privacy protection measures---a common phenomenon in security and privacy compliance \cite{wong2023privacy, forbestechcouncil2018gdpr}. Refer to the codebook in Table \ref{tab:psd_codebook} for the full list of supporting documentation types and their definitions.

\subsection{Proposed Protection Mechanisms}
We next turn to the tools and methods within DPG candidates' responses. 
Roughly, the privacy protection strategies they proposed can be categorized as `data-oriented' strategies (e.g., data strategies, secure data storage) and `process-oriented' strategies (e.g., governance processes/audits, vulnerability testing). Data-oriented strategies are technical privacy measures that directly operate on data \cite{hoepman2018privacy}. Process-oriented strategies, on the other hand, are organizational procedures that ensure responsible handling of data \cite{hoepman2018privacy}.
The full list of privacy protection mechanisms and their definitions is provided in Table \ref{tab:ppg_codebook}.

As shown in Figure \ref{fig:ppg}, the most common privacy protection mechanisms claimed by DPGs were data strategies, secure data storage, and access control -- around 26\% of the DPGs mention using mechanisms that fall under one or more of these categories. DPGs that use data strategies mention techniques such as minimizing the amount of data collected and anonymizing any personal data collected. Secure data storage mechanisms involve using measures like encryption to protect personal data. Mechanisms under access control enforce restrictions for accessing personal data based on predefined rules and policies.

Nearly 9\% of the DPGs propose taking responsibility for establishing and/or adhering to a governance process to ensure the protection of personal data. About 6\% of DPGs proposed to implement user controls to let users express privacy preferences effectively (e.g., provide consent, submit data deletion requests). Less common strategies  (1 - 3\%) include routinely applying security upgrades and patches, and performing vulnerability testing to ensure user data is protected. 
 
\subsection{Privacy Component Analysis}
Our final theme conducted a privacy coverage analysis, which was meant to understand whether DPGs are addressing common privacy considerations that arise in existing evaluation frameworks and guidelines. Since there is no single globally-adopted privacy framework or guideline, we extracted common components from five widely-used privacy frameworks and principles: the NIST Privacy Framework \cite{nist2024privacyframework}, LINDDUN \cite{wuyts2015linddun}, APEC Information Privacy Principles \cite{cooperation2005apec}, Privacy By Design Principles \cite{cavoukian2009privacy, hoepman2018privacy}, and principles outlined by the GDPR under Article 5 \cite{gdpr_art5}. 
We assigned one code to each concept or idea that appears in \emph{all} of the above resources, resulting in eight common components, listed in Figure \ref{fig:enter-label}. The definitions of these common components are provided in Table \ref{tab:pca_codebook} (Appendix \ref{app:codebooks}). 

In our analysis, we checked whether the DPGs addressed each of these components. We define `addressing a component' as including some amount of documentation describing their efforts related to that component. Results from our analysis indicate that most of the components do not achieve a high coverage rate. 

We find that \textbf{coverage of these common privacy components is sparse at best}.
The component with the highest coverage rate across DPGs was security safeguards (55\%), followed by regulatory efforts (33\%). Most DPGs that addressed the security safeguards component employed measures like access control and encryption to protect user information. DPGs that addressed the regulatory efforts component mentioned their efforts to comply with regulations like GDPR, and some DPGs mentioned the use of audits to review their compliance to these regulations. These results reinforce our earlier observation that some DPGs equate privacy protection with security safeguards and compliance efforts. For example:

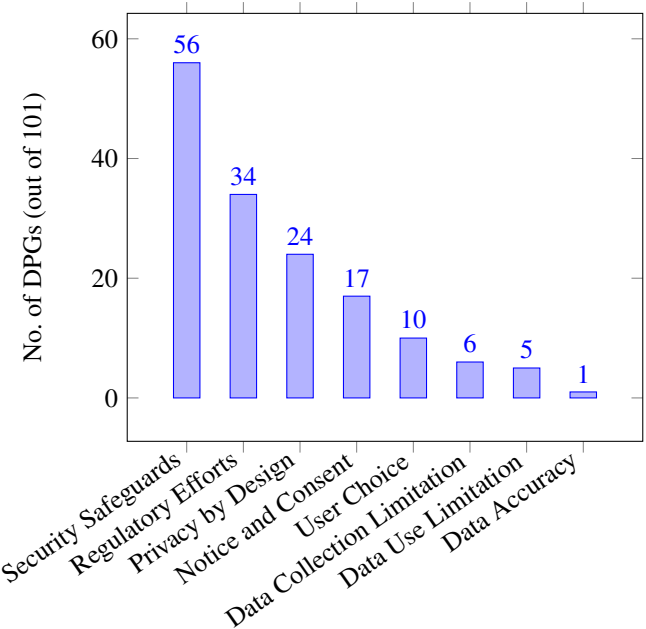
\begin{figure}
    \centering
    \begin{tikzpicture}
\begin{axis}[
ybar,
enlargelimits=0.15,
legend style={at={(0.5,-0.2)},
anchor=north,legend columns=-1},
ylabel={No. of DPGs (out of 101)},
symbolic x coords={
Security Safeguards,
Regulatory Efforts,
Privacy by Design,
Notice and Consent,
User Choice,
% Challenge Accuracy,
Data Collection Limitation,
Data Use Limitation,
% Privacy by Default,
Data Accuracy,
% Accountability,
},
xtick=data,
nodes near coords,
nodes near coords align={vertical},
x tick label style={rotate=35,anchor=east},
]
\addplot coordinates {
(Security Safeguards, 56)
(Regulatory Efforts, 34)
(Privacy by Design, 24)
(Notice and Consent, 17)
(User Choice, 10) % combined user choice and challenge accuracy
% (Challenge Accuracy, 5)
(Data Collection Limitation, 6)
(Data Use Limitation, 5)
% (Privacy by Default, 5) % combined this with privacy by design
(Data Accuracy, 1) 
% (Accountability, 0) % combined regulatory and accountability
};
\end{axis}
\end{tikzpicture}
    \caption{Results from qualitative analysis of DPG responses for Privacy Component Analysis.}
    \label{fig:pca}
    \vspace{-0.2in}
\end{figure}

\begin{quote}
    \textbf{DPG18}: ``... prioritizes security, stability, and scalability above all else, and many of our users implement ... to comply with GDPR, HIPAA, and other policies.''
\end{quote}

\begin{quote}
    \textbf{DPG21}: ``We ensure the security of collected data both by technical and policy means ... the number of personnel who have access to such data is strictly limited ... all personnel have signed non-disclosure agreements (NDAs), which clearly define user’s personal data as confidential information subject to confidentiality terms. The NDAs also imply strict monetary penalties in case of a breach. As to the technical level of ensuring security of the data, we use SSL certificates, database connections are private -> connection to the DB is available only via local server (outbound connections are disabled).''
\end{quote}
 
We also find that DPGs largely did not on address user notice, choice, and control. Only 17\% of the DPGs included documentation on notice and consent. Most DPGs that addressed this component mentioned that information on their data practices, such as purpose of data collection or third-party sharing, could be found in their privacy policies.

Furthermore, only around 10\% of the DPGs addressed user choice in their responses to the questionnaire. These DPGs provided their users with some amount of control in how their data is collected, used, or shared. For instance:
\begin{quote}
    \textbf{DPG54}: ``The demographic data (birth year and gender) fields are optional, and are not prerequisites for using the platform, allowing users for whom this information is more sensitive to opt out.''
\end{quote}

\begin{quote}
    \textbf{DPG75}: ``All information is transferred securely using HTTPS and raw data provided by the user for analysis can be deleted at the user's request.''
\end{quote}
Few DPGs allowed users to delete their data upon request.

\section{Case Studies of Digital Public Goods}
\label{sec:case_studies}

To gain a deeper understanding of the actual privacy practices and strategies employed by DPGs, we conducted an in-depth case study of three different DPGs. The case-studies allowed us to determine whether DPGs may have possible privacy implications that are not captured by the DPG standard. 
\update{\textbf{We re-emphasize that our findings are based only on documentation, and do not necessarily mean that the implementations have privacy vulnerabilities.}}

\subsection{Case Study 1: MOSIP}
The Modular Open Source Identity Platform (MOSIP) is an open-source and open standards foundational identity platform \cite{mosip}. It serves as an API-first platform for governments to build their own national ID platforms, offering ID life-cycle management and identity verification capabilities. 
The platform has over 100 million registered users and is operational in 11 countries, including Morocco, Ethiopia, and Sri Lanka.

\paragraph{Summary of Findings} MOSIP's response to Section 9(a)(iii) of the DPG questionnaire states that ``privacy and security practices are central to MOSIP and the project has taken extensive measures to provide security of data and has numerous existing and evolving features on privacy and data protection.'' MOSIP's response includes a link to its adopter-focused security and privacy documentation, which outlines the access control (e.g., authentication, rate-limiting) and secure data storage (e.g., encryption at rest) measures it supports. 

The MOSIP response and documentation was among the more careful of the DPGs we analyzed. At the same time, our threat elicitation process revealed potential issues, such as data being revealed to third parties in plaintext during authentication and secure storage not being used at all stages of data ingestion. MOSIP reported that most of our findings were mistakes in the documentation (not in the underlying software), some of which have since been updated. Nonetheless, the information collected by the DPG standard is not  nuanced enough to reveal such potential privacy vulnerabilities.

\subsubsection{Analysis}
The high-level architecture of MOSIP consists of two core modules: (1) ID Lifecycle Management, and (2) Authentication. ID lifecycle management includes several sub-components such as ID pre-registration, enrollment, updation and de-activation \cite{mosipdocs2024overview}. The authentication module provides ID authentication services \cite{mosipdocs2024idauthentication}. The data flow diagram (DFD) for MOSIP is illustrated in Figure \ref{fig:MOSIP_DFD} in the Appendix. This illustration informed the threat elicitation process using the LINDDUN framework \cite{wuyts2015linddun}.

Residents have the option to pre-register online and then visit designated centers to complete the registration process \cite{mosipdocs2024collabpre}. According to MOSIP's responses to the DPG questionnaire, the ID creation process needs residents to submit their legal name, age, address, biometrics (e.g., fingerprint, face, iris), and other PII as required by the country. When residents need to authenticate themselves with relying parties, these institutions serve as proxies to verify the residents digital IDs against MOSIP's servers \cite{mosipdocs2024idauthentication}.

\paragraph{Observation 1: Passing Clear Text Credentials to Relying Parties}
Authenticating a digital ID in MOSIP involves a relying party acting as a proxy to transmit credentials on behalf of the end-user. The relying party collects \textit{unencrypted} end-user 
\newrevision{virtual IDs (VIDs) and one-time passwords (OTPs)} and submits them to MOSIP's servers for verification. For this purpose, MOSIP utilizes a ``yes/no'' API to deliver verification responses and places trust in relying parties that may belong to private or government organizations \cite{mosipdocs2024idauthentication}. 

However, over-reliance on these parties can lead to the misuse of user credentials received in clear text, allowing them to identify users even when temporary VIDs are used.
This poses a privacy risk as the clear text credentials could be intercepted, compromising the identity and personal information of users \update{as noted in prior work, which found a related  vulnerability in OAuth 2.0 implementations \cite{oauth}}.
\newrevision{Note that MOSIP offers an alternative authentication mechanism called eSignet, which mitigates this risk.}

\paragraph{Observation 2: Weak Anonymization in Profiling System}
MOSIP offers an `Anonymous Profiling System' \cite{mosipdocs2024anonymousprofiling} for conducting privacy-preserving analytics on pre-registration data. The anonymized dataset \cite{mosipdocs2024anonymousprofiling} includes personal information attributes like gender, location, and year of birth. Documentation indicates that anonymization is provided through suppression of data. Suppression is a form of weak anonymization that could introduce potential privacy vulnerabilities, as malicious actors may carry out reconstruction attacks \cite{dwork2017exposed} by launching targeted queries against the profiling system.
\newrevision{MOSIP counters the risk of linkage attacks by encrypting the database so that a record is unidentifiable without knowledge of the corresponding VID. Nonetheless, depending on which fields are shared with third parties, inference attacks using correlated data sources have been used in other contexts to de-anonymize users based on partial information \cite{sweeney2000simple,narayanan2008robust,cohen2022attacks}, as well as inferring properties of groups of users. }

\paragraph{Observation 3: Unencrypted Storage of Pre-Registration Data}
MOSIP's `pre-registration' databases are downloaded to an operator's system for offline data retrieval \cite{mosipdocs2024anonymousprofiling}. However, \newrevision{in August 2023, at the time this case study was conducted,} the documentation  suggested that these databases are stored in an unencrypted format, without providing a justification for doing so \cite{mosipdocs2024anonymousprofiling}. 
\newrevision{The documentation has been updated since we shared our findings with MOSIP in September 2023; as of June 2024, it states that pre-registration data is indeed stored in encrypted form}.

\paragraph{Observation 4: Unclear Documentation of Data Retention and Deletion Policies}
The documentation on MOSIP's data retention and deletion policies is unclear, as it uses two different terms: `deactivation' and `decommission'. Deactivation refers to temporary shutdown, while decommission refers to permanent shutdown of a resource \cite{mosipdocs2024administration}. It is unclear which option (if any) leads to permanent deletion of user data, including biometrics. 

\paragraph{Observation 5: Possibly Low-Quality Informed Consent}
During registration, the operator can choose to mark consent on behalf of the individual \cite{mosipdocs2024registrationclient}. This raises concerns about the quality of \update{informed consent \cite{friedman2000informed}}, as operators could mark consent without clearly explaining the terms to individuals. 

\paragraph{Responsible Disclosure}
\newrevision{We communicated our observations with MOSIP, who confirmed that Observation 3 was a documentation gap. MOSIP has since updated that documentation, and more generally,  significantly clarified their documentation of privacy data flows compared to when we ran this study.} 

\subsection{Case Study 2: Ushahidi}
Ushahidi is a crowd-sourcing platform for social activism. It aims to map and document information during political campaigns, natural disasters, and other events of public interest \cite{ushahidi}. The platform enables local observers to easily submit reports via their mobile phones or the internet, creating an archive of events accompanied by geographic and time-date details.
Ushahidi has been deployed in over 60 countries and supports more than 40 languages 
Some of its use cases include supporting earthquake relief efforts in Nepal, ensuring fair elections in Nigeria, and helping women address sexual violence in Egypt \cite{ushahidi_deployments}.

\paragraph{Summary of Findings}
Ushahidi's response to Section 9(a)(iii) of the DPG questionnaire consists of links to their documentation on how their platform supports data security and the measures implementers must take to comply with GDPR. Specifically, their response describes ``reasonable administrative, physical and electronic measures'' like encrypting data in transit, securing servers using access control mechanisms like (i) restricting open ports, (ii) using hardened SSL configurations, and limiting communication between services to internal private networks.
Ushahidi also provides implementers with instructions on collecting consent from users. 

However, their response does not provide details about privacy measures (e.g., anonymization) for PII collected from sources other than surveys (e.g., Twitter, emails). Of concern, this data may still be stored on the platform even after the original data sources have been deleted. Moreover, their response states they use security safeguards during transit, but whether they encrypt this data at rest is unclear. 
Note that in the time since this study was conducted (Fall 2023), we have noticed a dramatic (possibly total) drop in Ushahidi deployments' posts sourced from X (formerly Twitter).

\subsubsection{Analysis}
Ushahidi's high-level architecture consists of three core components: the Platform, Services, and Data. The `Platform' component includes Ushahidi's core platform and MySQL data store \cite{ushahidi_tech_stack}. The `Services' component provides POST and REST APIs for ingesting data and transmitting reports to Ushahidi's web interface and mobile application app \cite{ushahidi_tech_stack}. According to Ushahidi's responses to the DPG questionnaire, the platform can collect ``email addresses, location, and telephone numbers'' of its users.

The `Data' component allows implementers to configure input data sources \cite{ushahidi_tech_stack}. End-users can submit reports via Ushahidi's web interface or send reports to dedicated email or SMS channels. Additionally, the platform can be integrated with Twitter (now known as X) to ingest data based on hashtags \cite{ushahidi_twitter}. The data flow diagram (DFD) for Ushahidi is illustrated in Figure \ref{fig:Ushahidi_DFD}. 

\paragraph{Observation 1: Inconsistent Data Updates}
Ushahidi supports the use of Twitter’s (now known as X) developer API to collect messages (or tweets) based on hashtags \cite{ushahidi_twitter}. This functionality aids in monitoring crisis response, elections, political and community engagements. The content from collected tweets is stored in a database called ‘messages'. It is observed that content from deleted or modified tweets does not get updated on Ushahidi’s platform \cite{ushahidi_twitter}. As a result, data stored on the platform may become outdated and no longer reflective of the current state of affairs when real-time updates are not received.
This relates to privacy and data use because users could choose to remove content on Twitter, but have it remain active on Ushahidi. \update{This counters user privacy expectations around data deletion \cite{minaei2022empirical}.}

\paragraph{Observation 2: Limited Anonymization Coverage}
Ushahidi aggregates data from various sources, including user-submitted reports (surveys), Twitter, email, and SMS \cite{ushahidi_datasources}. The platform provides an optional anonymization control that allows platform administrators to selectively obfuscate an author's information, location and timestamps \cite{ushahidi_data_obfuscation}. 

From the documentation, it is unclear whether data anonymization features are available for information collected from sources other than Ushahidi Surveys \cite{ushahidi_surveys}. 
The possible lack of anonymization features for other sources could pose a privacy risk to the reporter's identity.

\paragraph{Observation 3: Lack of Privacy Safeguards for Raw Data}
The Ushahidi platform offers anonymization features for publishing posts to end-users. Platform admins can optionally choose to obfuscate an display fields such as author’s information, location and timestamps \cite{ushahidi_surveys}.
However, the data is stored in plain text in the database without employing any data-oriented strategies (e.g., anonymization, obfuscation) to protect privacy. Storing plain-text data in the database could pose a risk \cite{pearson2024breach} to reporters' privacy in certain contexts where trust is assumed: (1) malicious administrators with access to internal databases, or (2) raw data shared for secondary purposes such as research, policy-making, or compliance with law enforcement requests.

\paragraph{Observation 4: Use of Direct Identifiers for Unstructured Data Sources}
The collected data contain direct identifiers, such as the author’s information. 
Structured data from in-platform surveys are obfuscated, while data from unstructured sources (such as email, SMS and Twitter reports) are stored and/or published without applying anonymization techniques.
The use of direct identifiers in a crowdsourcing platform could single out and identify the reporter who submitted the information. Depending on the context, this may pose a serious risk or threat to the reporter (e.g., activist campaigns). Additionally, reporters' unique identifiers can be used to correlate with social networks to discover personal associations, posing a serious risk or threat not only to the reporter but also their close connections (e.g., friends or family).

\paragraph{Responsible Disclosure} We have shared our observations with Ushahidi's security team on 12/11/23, but we have not received a response at the time of publication.

\subsection{Case Study 3: DIVOC}
The Digital Infrastructure for Verifiable Open Credentialing (DIVOC) is an open-source platform for countries to conduct large-scale digitized health campaigns \cite{divoc2024homepage}. Adopters can flexibly choose the components they want to implement and customize them to suit their needs. For example, countries can use DIVOC to establish a digital infrastructure for issuing and verifying their citizens' vaccination certificates. 

DIVOC is developed and maintained by the eGov Foundation of India. It has been used by countries like India, Indonesia, Jamaica, the Philippines, and Sri Lanka to issue and verify over 2 billion COVID-19 vaccination certificates \cite{divoc2024homepage}.

\paragraph{Summary of Findings}
DIVOC's response to Section 9(a) of the DPG questionnaire states that they do not collect PII. However, their infrastructure allows implementers to collect PII while orchestrating health campaigns. We observe that the DPG standard is not nuanced enough to differentiate between the collection PII by the DPG or its implementers. For example, the other two DPGs we evaluated (MOSIP and Ushahidi) are used by implementers who collect PII, but the DPGs still declare they collect PII in their responses. Although DIVOC mentions implementers are responsible for protecting user privacy in their response, they also provide privacy and security best practices  in their adopter-focused documentation. 

\subsubsection{Analysis}
The DIVOC platform follows a microservice architecture and can integrate with third-party services \cite{divoc2024architecture}. The DFD for DIVOC is illustrated in Figure \ref{fig:DIVOC_DFD} in the Appendix. This illustration informed the threat elicitation process using the LINDDUN framework \cite{wuyts2015linddun}. At a high level, DIVOC consists of two core modules \cite{divoc2024modules}. The first module is responsible for issuing, verifying, and distributing credentials (e.g., vaccination certificates). The second module monitors the performance of the health campaign by computing real-time analytics. 

Countries can include several additional modules \cite{divoc2024modules} in their DIVOC instance, such as a program set-up module that creates and maintains registries for credentials and facilities where these credentials are issued. 
A citizen portal is also available for citizens to self-register, schedule appointments with a facility, and download and verify their credentials. 

\paragraph{Observation 1: Delegation of Responsibilities}
DIVOC states that they do not collect, store, or distribute PII in their response to the DPG questionnaire. However, their platform is ``meant for last-mile vaccination administration and credentialing'', and its implementers can collect and store PII such as name, date of birth, and identifiers like a national identity number \cite{divoc2024collectedpii}. DIVOC mentions their platform architecture prioritizes data minimalism, with ``well-designed privacy \& security'' measures in their response. They further note that the effectiveness of the supported privacy measures depends on the individual privacy policies used by their adopters. Although they delegate the responsibility of protecting user privacy to their adopters, DIVOC provides them with privacy and security best practices to follow \cite{divoc2024platformpolicy}, described below. 

\paragraph{Observation 2: Privacy Guidelines for Adopters}
DIVOC provides comprehensive data protection guidelines for its adopters in its documentation \cite{divoc2024platformpolicy}. For example, to ensure secure data backups, DIVOC recommends implementing the principle of least privilege by restricting access to user and system information based on task requirements, as well as purging intermediate data backups and keeping full backups on separate servers after encrypting the data. 
It also gives recommendations on authentication and password management, access control, and platform updates. 
Finally, DIVOC also includes templates of user-facing privacy policies that adopters can use while running their health campaigns \cite{divoc2024shortprivpolicy, divoc2024longprivpolicy}. 

\paragraph{Responsible Disclosure} We had no privacy concerns to share. % to the parent DPG.

\section{Discussion}
Our findings highlight three important points: 

\noindent \textbf{1. The DPG standard is not currently ensuring that DPGs offer a strong level of privacy protection.}
Although the intent of the privacy question on the DPG questionnaire is clearly aligned with best privacy practices, the reality is that many approved DPGs have responded to it incompletely or incorrectly, and made it through the approval process. 
For example, our qualitative analysis in Section \ref{sec:qual_analysis} of DPG responses indicates that over 65\% of DPGs we studied either had incomplete or vague privacy documentation; if these responses are representative of their true privacy posture, those DPGs may be vulnerable to privacy threats. 
Hence, the DPG questionnaire is not currently filtering out responses with a weak or incomplete description of privacy protections. 

\smallskip
\noindent \textbf{2. The current DPG standard does not collect nuanced enough information to distinguish DPGs with very different privacy profiles.}
Among certified DPGs, there is a broad range of levels of privacy maturity. 
For instance, we noted that MOSIP had implemented and documented many privacy features, whereas DIVOC chose to implement relatively fewer privacy features, leaving a significant amount of implementation to the DPG adopter.
There could exist a version of the DPG standard that differentiates between these two very different models of implementation. We give one proposal for how to design such a model in Section \ref{sec:recs}. 
However, we note that the DPGA may not wish to be responsible for  differentiating between DPGs of differing privacy postures, as this would require a much more in-depth analysis. 

\smallskip
\noindent \textbf{3. Should the DPGA be evaluating DPGs' privacy posture? } A broader question is whether the DPGA should be tasked with evaluating or ensuring the privacy of DPGs. 
Currently, the DPGA may be constrained in part by the UN's report, which emphasizes the importance of  privacy, and in part by the lack of clear guidance internationally on how to evaluate the privacy of software (let alone other classes of DPGs like machine learning models, datasets, etc.). Hence, it may be worth revisiting whether the DPGA's role in privacy evaluation. We discuss an alternative model in Section \ref{sec:recs}.

\subsection{Limitations} 
\label{sec:limitations}
Our methodology has some limitations, which we highlight here. 
First, our DPG sample is biased, including only DPGs that were certified. 
It would be useful to also analyze the responses of DPG candidates that were not approved, but this data is not publicly available. 
% For instance, Glific is a DPG that allows implementers to create WhatsApp chatbots and send personalized messages to multiple contact lists \cite{glific2024homepage}. Glific claims not to collect, store, or distribute PII in its DPG questionnaire response. However, its implementers must demonstrate to Glific that they have obtained consent to send WhatsApp messages from each individual in their contact lists \cite{glific2024onboardingtoolkit}. It is unclear whether Glific retains this information from their response to the DPG questionnaire.

Another important limitation of our methodology is that it  rewards DPGs with more developed privacy documentation, regardless of how developed their privacy features are. 
For example, as we saw in our case studies in Section \ref{sec:case_studies}, DIVOC fared well in part because it did not specify many implementation details for privacy functionalities. Instead, it delegated responsibility to solution adopters, and documented recommendations clearly in its documentation. 
This prevented our threat elicitation process from identifying threats in its data flows. On the other hand, MOSIP implemented (and documented) more privacy features, so it was easier to observe concrete gaps. A more complete prototype like MOSIP may require less effort from adopters, who will most likely use out-of-the-box privacy features. However, it is difficult to directly compare DPGs with differing levels of implementation.

\subsection{Recommendations}
\label{sec:recs}

\subsubsection{DPG Community}
\paragraph{(1) Do not refine the DPG questionnaire with more specific questions.}
\update{A natural reaction to our findings is to attempt to revise the DPG standard to be more precise and granular about what privacy properties a DPG should satisfy. We suggest \emph{not} pursuing such a direction. Since the DPG standard is meant to be adopted globally, building consensus around a privacy standard is likely to be politically challenging. Privacy norms are highly culture-specific \cite{zabihzadeh2019cultural}, and we note that to the best of our knowledge, there are no true privacy standards in place that address an entire product, even among (inter-)national standards bodies; instead, the focus has been on building general-purpose \emph{frameworks} that are very high-level, but also broadly applicable \cite{privacyref2024framework}.
Second, privacy best practices are often technology-specific, and it is unclear how to craft a standard that  encompasses the broad range DPGs (e.g., a national ID system vs. a machine learning model). }

\begin{figure}[htp]
    \centering
    \includegraphics[width=0.45\textwidth]{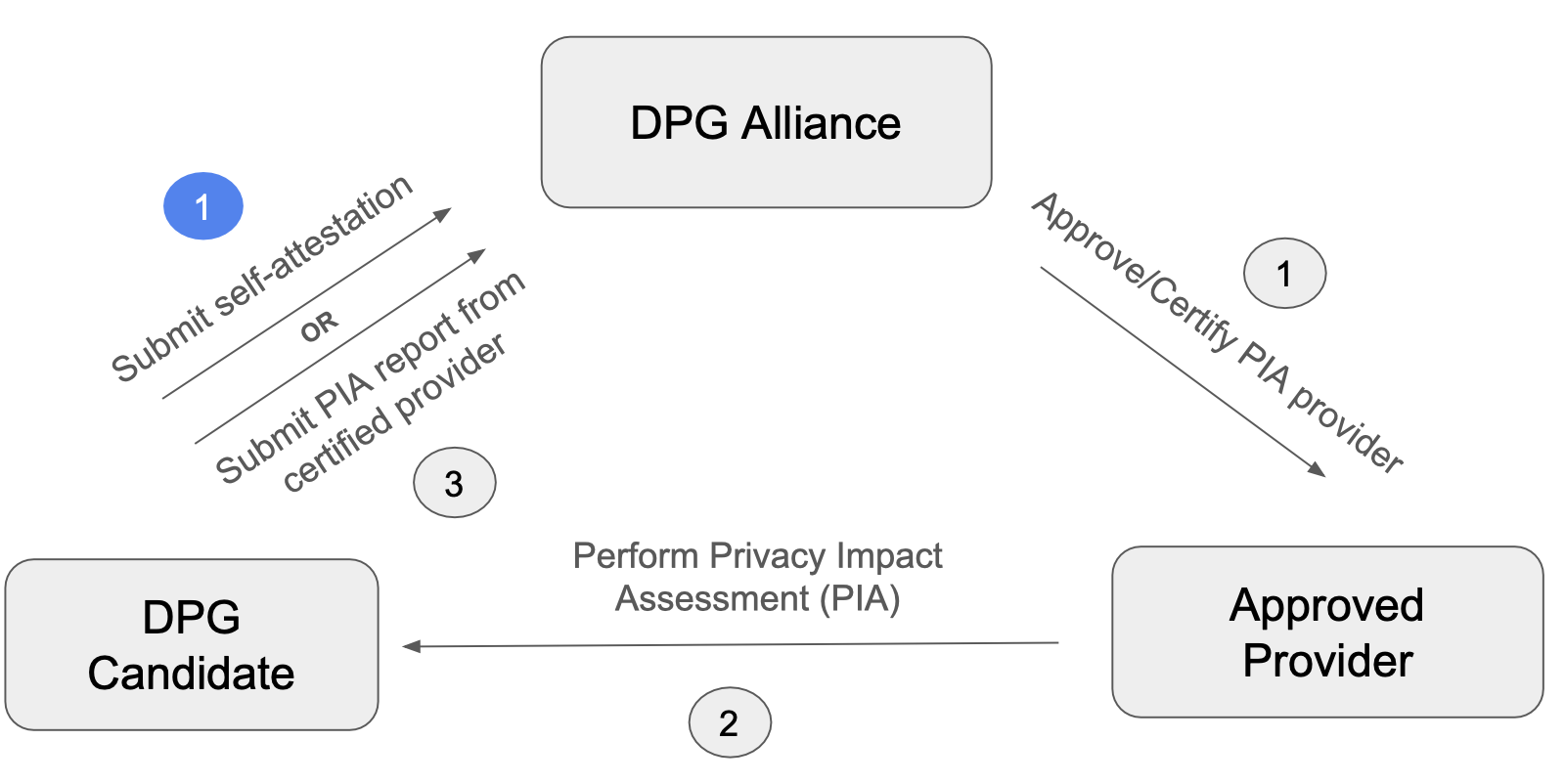}
    \caption{3-stakeholder model to facilitate DPG privacy evaluation. The third-party assessment would involve the gray sequence of steps, whereas a self-assessment would require only the single blue step. 
    }
    \label{fig:stakeholder_interaction}
\end{figure}

\paragraph{(2) Adopt a new architecture for collecting privacy evaluations.}
Instead of updating the DPG standard to be more comprehensive, we suggest a model that makes use of the existing ecosystem for privacy evaluation, which are themselves the products of many years of refinement and stakeholder engagement \cite{nist2024privacyframework,wuyts2015linddun,PIA}. 
Our proposed model would have two tiers of privacy certification (see Figure \ref{fig:stakeholder_interaction}).

\emph{Tier 1: Certified Privacy Impact Assessment.}
At the stronger tier, the DPGA would ask candidate DPGs to submit documentation attesting to the fact that they underwent a Privacy Impact Assessment (PIA) \update{or a comparable regional variant, such as the Singaporean Data Protection Trustmark \cite{dptm}} from a certified provider.
A PIA is an analysis of how personally identifiable information is collected, used, shared, and maintained; it involves answering a list of questions regarding data collection, retention, use, and more \cite{PIA}. 
It provides a more fine-grained view of a product's privacy posture than the current DPG questionnaire, as studied in this work. 
PIAs have international adoption and are currently mandated for U.S. federal agencies by the e-Commerce Act of 2002 \cite{ecommerce} and by the E.U. through GDPR Article 35 for all high-risk data processing activities \cite{gdpr-article35}.
Under our suggested process, the DPGA would collect and publish evidence of a PIA (or a comparable alternative) from an approved provider; \update{the DPGA can maintain a list of acceptable assessment tools and assessors e.g., \cite{onetrust,osano}}. 
In addition, DPGs would upload the outcomes from the audit (answers to all questions), which should be made available on the DPGA website. 

\emph{Tier 2: Self-Assessment.}
In the second tier, DPG candidates would submit a self-attestation that they underwent a PIA. The documentation from that process would be uploaded along with the candidate's self-attestation, so potential users can view the DPG's self-evaluated privacy posture. 

Under our proposed model, the DPGA would approve DPGs as long as they have accomplished one of the two. 
In particular, the DPGA would \emph{not} directly evaluate, or provide their seal of approval, to DPGs' privacy postures. 
Evaluation of privacy documentation would be handled by the adopting entity. 
Note that many DPGs are not hosted services, but require an integrator (often a government) to host and run the DPG. 
In these cases, it is reasonable to expect a government to expend  resources to evaluate the privacy posture of a piece of software before using it on constituents' data. 

A potential drawback of this suggested architecture is that it increases the barrier to entry for new DPGs. 
However, privacy was presented in the UN's mandate as a first-order requirement of DPGs. If this is the case, 
it may be necessary to raise the barrier to DPG certification to ensure that DPGs are handling user data properly. 
\update{We provide a stakeholder cost analysis comparing the two tiers in Appendix \ref{app:cost_analysis}. }

\subsubsection{Research Community}
\paragraph{(1) Further research is needed on communicating the privacy posture of DPGs to downstream adopters.}
\update{There is an active body of research on communicating the privacy posture of applications to end users \cite{hagen2016user,kelley2009nutrition,zimmeck2024generalizable}. One well-known example is privacy nutrition labels \cite{kelley2009nutrition,li2022understanding}. 
These technologies may be nontrivial to apply to DPGs. 
Adopters of a DPG could be governments or hobbyists, and they may use the same tool for very different purposes. 
Hence, their privacy needs may vary significantly, so the structure that makes privacy nutrition labels easy for users to understand may not extend easily to DPGs. 
Second, DPGs often limit what aspects of the system they implement, and which parts they leave to the downstream adopter. This can impact privacy in nuanced ways (as shown in our case studies in \S\ref{sec:case_studies}), and those impacts should be communicated clearly.
In sum, understanding how to clearly communicate the privacy (and security) posture of a DPG is an interesting and complex question for the usable security and privacy research community. }

\paragraph{(2) Continue to develop automated tools for dynamically evaluating the privacy posture of software.} 
A drawback of the suggested architecture is staleness; a privacy audit typically has a short shelf life because every new feature can introduce new privacy vulnerabilities. Hence, inspired by OSCAL \cite{nist_oscal}, continued research is needed on automatically processing a codebase and extracting potential privacy vulnerabilities. 
While this is already a rich area of research \cite{reuben2016automated, accorsi2008automated,kunz2023privacy}, there is still room to make these tools usable and connect them to standardized privacy certifications. 

\subsection{Ethical Considerations}
This study is not human subjects research, and it used only public data about products (not people). The study was not subject to review by our Internal Review Board. We followed industry-standard best practices for disclosing potential  vulnerabilities to MOSIP and Ushahidi after our case studies, and gave both  60 days' notice prior to publicizing results \cite{cisa}.

\section{Conclusion}
This work provides the first large-scale study of DPGs and their privacy properties. 
Our results suggest that the DPG standard may benefit from revising its methodology for evaluating DPG candidates' privacy maturity. 
We have communicated our findings and recommendations with the DPGA, which is currently  revising the DPG standard (although we are not sure in which direction).
\update{In addition to encouraging the DPGA to improve the DPG privacy certification process, we hope this study will inform future privacy-conscious initiatives, such as the emerging push for digital public infrastructure \cite{DPI}. We also hope this work will encourage the usable privacy research community to explore ways of communicating privacy to third-party adopters of user-facing technologies.}

\ifpublicversion
\section*{Acknowledgments}
This work was made possible by the Bill \& Melinda Gates Foundation. The views expressed in this work are solely our own. We would like to acknowledge valuable feedback from Assane Gueye, Lorrie Cranor, Lujo Bauer, MOSIP, and the DPGA, as well as the anonymous shepherd and reviewers of this work. 
\fi

\bibliographystyle{plain}
\bibliography{ref}

\begin{thebibliography}{10}

\bibitem{onezero2024tech}
Tech monopolies and the insufficient necessity of interoperability.
\newblock
  \url{https://onezero.medium.com/tech-monopolies-and-the-insufficient-necessity-of-interoperability-aafba94f1eb3},
  2024.
\newblock (Accessed on February 14, 2024).

\bibitem{accorsi2008automated}
Rafael Accorsi.
\newblock Automated privacy audits to complement the notion of control for
  identity management.
\newblock In {\em Policies and Research in Identity Management: First IFIP
  WG11. 6 Working Conference on Policies and Research in Identity Management
  (IDMAN'07), RSM Erasmus University, Rotterdam, The Netherlands, October
  11-12, 2007}, pages 39--48. Springer, 2008.

\bibitem{act1996health}
Accountability Act.
\newblock Health insurance portability and accountability act of 1996.
\newblock {\em Public law}, 104:191, 1996.

\bibitem{adjerid2016impact}
Idris Adjerid, Alessandro Acquisti, Rahul Telang, Rema Padman, and Julia
  Adler-Milstein.
\newblock The impact of privacy regulation and technology incentives: The case
  of health information exchanges.
\newblock {\em Management Science}, 62(4):1042--1063, 2016.

\bibitem{dgpa2024registry}
Digital Public~Goods Alliance.
\newblock Digital public goods registry.
\newblock \url{https://digitalpublicgoods.net/registry/}.

\bibitem{dpga2024standard}
Digital Public~Goods Alliance.
\newblock Digital public goods standard questionnaire.
\newblock
  \url{https://github.com/DPGAlliance/DPG-Standard/blob/main/standard-questions.md}.

\bibitem{ashley2002privacy}
Paul Ashley, Calvin Powers, and Matthias Schunter.
\newblock From privacy promises to privacy management: a new approach for
  enforcing privacy throughout an enterprise.
\newblock In {\em Proceedings of the 2002 workshop on New security paradigms},
  pages 43--50, 2002.

\bibitem{dptm}
Infocomm Media~Development Authority.
\newblock Data protection trustmark (dptm) certification.
\newblock
  \url{https://www.imda.gov.sg/how-we-can-help/data-protection-trustmark-certification}.
\newblock (Accessed on May 23, 2024).

\bibitem{forbes2021trends}
Peter Bendor-Samuel.
\newblock Trends in rising prices affecting companies using third-party
  services.
\newblock
  \url{https://www.forbes.com/sites/peterbendorsamuel/2021/08/02/trends-in-rising-prices-affecting-companies-using-third-party-services/?sh=2977dc4827b2},
  August 2 2021.
\newblock (Accessed on February 14, 2024).

\bibitem{cavoukian2009privacy}
Ann Cavoukian.
\newblock Privacy by design.
\newblock 2009.

\bibitem{chen2023motivating}
Yan Chen, Rosta Farzan, Robert Kraut, Iman Yeckehzaare, and Ark~Fangzhou Zhang.
\newblock Motivating experts to contribute to digital public goods: A
  personalized field experiment on wikipedia.
\newblock {\em Management Science}, 2023.

\bibitem{cohen2022attacks}
Aloni Cohen.
\newblock Attacks on deidentification's defenses.
\newblock In {\em 31st USENIX Security Symposium (USENIX Security 22)}, pages
  1469--1486, 2022.

\bibitem{cooperation2005apec}
Asia-Pacific~Economic Cooperation.
\newblock Apec privacy framework.
\newblock {\em Asia Pacific Economic Cooperation Secretariat}, 81, 2005.

\bibitem{forbestechcouncil2018gdpr}
Forbes~Tech Council.
\newblock Gdpr and the 'security by compliance' mistake.
\newblock
  \url{https://www.forbes.com/sites/forbestechcouncil/2018/07/02/gdpr-and-the-security-by-compliance-mistake/?sh=4d0d78fecc4d},
  July 2 2018.
\newblock (Accessed on February 14, 2024).

\bibitem{cisa}
{Cybersecurity \& Infrastructure Security Agency}.
\newblock Bod 20-01: Develop and publish a vulnerability disclosure policy |
  cisa.
\newblock
  \url{https://www.cisa.gov/news-events/directives/bod-20-01-develop-and-publish-vulnerability-disclosure-policy#:~:text=Many%20in%20the%20security%20research,the%20issue%20without%20unnecessary%20delay.}
\newblock (Accessed on February 09, 2024).

\bibitem{dpga}
{Digital Public Goods Alliance}.
\newblock Digital public goods.
\newblock \url{https://digitalpublicgoods.net/digital-public-goods/}.
\newblock (Accessed on February 14, 2024).

\bibitem{dpgstandard-overview}
{Digital Public Goods Alliance}.
\newblock Digital public goods standard.
\newblock \url{https://digitalpublicgoods.net/standard/}.
\newblock (Accessed on February 14, 2024).

\bibitem{dpgalliance2024repository}
{Digital Public Goods Alliance}.
\newblock Digital public goods repository.
\newblock
  \url{https://github.com/DPGAlliance/publicgoods-candidates/tree/main/digitalpublicgoods},
  2024.
\newblock (Accessed on February 14, 2024).

\bibitem{digitalpublicgoods}
{Digital Public Goods Alliance}.
\newblock Who we are.
\newblock \url{https://digitalpublicgoods.net/who-we-are/}, n.d.
\newblock (Accessed on February 14, 2024).

\bibitem{dwork2017exposed}
Cynthia Dwork, Adam Smith, Thomas Steinke, and Jonathan Ullman.
\newblock Exposed! a survey of attacks on private data.
\newblock {\em Annual Review of Statistics and Its Application}, 4:61--84,
  2017.

\bibitem{eaves2022best}
David Eaves, Leonie Bolte, Omayra Chuquihuara~Gozalo, and Surabhi
  Hodigere~Raghavendra.
\newblock Best practices for the governance of digital public goods.
\newblock {\em Ash Center Policy Briefs Series}, 2022.

\bibitem{divoc2024architecture}
eGov Foundation.
\newblock Divoc architecture.
\newblock \url{https://divoc.digit.org/platform/divoc-architecture}.

\bibitem{divoc2024modules}
eGov Foundation.
\newblock Divoc demo for modules.
\newblock \url{https://divoc.digit.org/divoc-demo}.

\bibitem{divoc2024homepage}
eGov Foundation.
\newblock Divoc egov foundation.
\newblock \url{https://divoc.egov.org.in/}.

\bibitem{divoc2024platformpolicy}
eGov Foundation.
\newblock Divoc platform policy guidelines.
\newblock
  \url{https://divoc.digit.org/community/about-project-team/platform-policy-guidelines}.

\bibitem{divoc2024longprivpolicy}
eGov Foundation.
\newblock Divoc privacy policy: Detailed.
\newblock
  \url{https://divoc.digit.org/community/about-project-team/privacy-policy-detailed}.

\bibitem{divoc2024shortprivpolicy}
eGov Foundation.
\newblock Divoc privacy policy: Short version for display.
\newblock
  \url{https://divoc.digit.org/community/about-project-team/privacy-policy-short-version-for-display}.

\bibitem{divoc2024collectedpii}
eGov Foundation.
\newblock Divoc: What information goes into a qr code?
\newblock
  \url{https://divoc.digit.org/platform/divocs-verifiable-certificate-features-2.0/what-information-goes-into-a-qr-code}.

\bibitem{worldbank2024lowincome}
Marcello Estevao.
\newblock 4 ways low-income economies can boost tax revenue without hurting
  growth.
\newblock
  \url{https://blogs.worldbank.org/voices/4-ways-low-income-economies-can-boost-tax-revenue-without-hurting-growth},
  2024.
\newblock (Accessed on February 14, 2024).

\bibitem{fedele2022dangerous}
Alessandro Fedele and Cristian Roner.
\newblock Dangerous games: A literature review on cybersecurity investments.
\newblock {\em Journal of Economic Surveys}, 36(1):157--187, 2022.

\bibitem{friedman2000informed}
Batya Friedman, Edward Felten, and Lynette~I Millett.
\newblock Informed consent online: A conceptual model and design principles.
\newblock {\em University of Washington Computer Science \& Engineering
  Technical Report 00--12--2}, 8, 2000.

\bibitem{level2dfds}
{GeeksforGeeks}.
\newblock Levels in data flow diagrams (dfd).
\newblock
  \url{https://www.geeksforgeeks.org/levels-in-data-flow-diagrams-dfd/}, 2024.
\newblock (Accessed on February 14, 2024).

\bibitem{un-mandate}
United Nations~Secretary General.
\newblock Roadmap for digital cooperation.
\newblock
  \url{https://www.un.org/en/content/digital-cooperation-roadmap/assets/pdf/Roadmap_for_Digital_Cooperation_EN.pdf},
  2020.
\newblock (Accessed on February 14, 2024).

\bibitem{gillwald2019governance}
Alison Gillwald and Anri van~der Spuy.
\newblock The governance of global digital public goods: Not just a crisis for
  africa.
\newblock {\em GigaNet, Berlin}, 2019.

\bibitem{aamdigital2024homepage}
Aam~Digital GmbH.
\newblock Aam digital.
\newblock \url{https://aam-digital.com/}.

\bibitem{ecommerce}
U.S. Government.
\newblock {Public Law 107–347, 107th Congress}.
\newblock
  \url{https://www.govinfo.gov/content/pkg/PLAW-107publ347/pdf/PLAW-107publ347.pdf},
  December 2022.
\newblock (Accessed on February 10, 2024).

\bibitem{gramm1999gramm}
Phil Gramm.
\newblock Gramm--leach--bliley act.
\newblock In {\em Vol. Public Law 106--102). Washington, DC: United States
  Congress}, 1999.

\bibitem{gray2007privacy}
Kendra Gray.
\newblock The privacy rule: Are we being deceived.
\newblock {\em DePaul J. Health Care L.}, 11:89, 2007.

\bibitem{hagen2016user}
Margaret Hagen.
\newblock $\{$User-Centered$\}$ privacy communication design.
\newblock In {\em SOUPS}, 2016.

\bibitem{hoepman2018privacy}
J-H Hoepman.
\newblock Privacy design strategies (the little blue book).
\newblock 2018.

\bibitem{microsoft2006uncover}
Michael Howard and David LeBlanc.
\newblock Uncover security design flaws using the stride approach.
\newblock {\em MSDN Magazine}, November 2006.
\newblock (Accessed on February 14, 2024).

\bibitem{gdpr_art5}
{Intersoft Consulting}.
\newblock Article 5 - principles relating to processing of personal data.
\newblock \url{https://gdpr-info.eu/art-5-gdpr/}, 2023.
\newblock (Accessed on November 10, 2023).

\bibitem{kelley2009nutrition}
Patrick~Gage Kelley, Joanna Bresee, Lorrie~Faith Cranor, and Robert~W Reeder.
\newblock A" nutrition label" for privacy.
\newblock In {\em Proceedings of the 5th Symposium on Usable Privacy and
  Security}, pages 1--12, 2009.

\bibitem{kelly2012investing}
Brian~B Kelly.
\newblock Investing in a centralized cybersecurity infrastructure: Why
  hacktivism can and should influence cybersecurity reform.
\newblock {\em BUL Rev.}, 92:1663, 2012.

\bibitem{kummer2020unemployment}
Michael Kummer, Olga Slivko, and Xiaoquan Zhang.
\newblock Unemployment and digital public goods contribution.
\newblock {\em Information Systems Research}, 31(3):801--819, 2020.

\bibitem{kunz2023privacy}
Immanuel Kunz, Konrad Weiss, Angelika Schneider, and Christian Banse.
\newblock Privacy property graph: Towards automated privacy threat modeling via
  static graph-based analysis.
\newblock {\em Proceedings on Privacy Enhancing Technologies}, 2023.

\bibitem{lam2019data}
WYNNE LAM and Bruce Lyons.
\newblock Data protection legislation and investment incentives when consumers
  are loss averse.
\newblock 2019.

\bibitem{lelarge2009economic}
Marc Lelarge and Jean Bolot.
\newblock Economic incentives to increase security in the internet: The case
  for insurance.
\newblock In {\em IEEE INFOCOM 2009}, pages 1494--1502. IEEE, 2009.

\bibitem{li2022understanding}
Tianshi Li, Kayla Reiman, Yuvraj Agarwal, Lorrie~Faith Cranor, and Jason~I
  Hong.
\newblock Understanding challenges for developers to create accurate privacy
  nutrition labels.
\newblock In {\em CHI}, 2022.

\bibitem{3rdpartyservices2018}
Emma L.~Slade Marijn~Janssen, Nripendra P.~Rana and Yogesh~K. Dwivedi.
\newblock Trustworthiness of digital government services: deriving a
  comprehensive theory through interpretive structural modelling.
\newblock {\em Public Management Review}, 20(5):647--671, 2018.

\bibitem{minaei2022empirical}
Mohsen Minaei, Mainack Mondal, and Aniket Kate.
\newblock Empirical understanding of deletion privacy: Experiences,
  expectations, and measures.
\newblock In {\em USENIX Security}, pages 3415--3432, 2022.

\bibitem{mosip}
{MOSIP}.
\newblock A digital public good for identity.
\newblock \url{https://mosip.io/#1}.
\newblock (Accessed on February 14, 2024).

\bibitem{mosipdocs2024administration}
{Mosip Documentation}.
\newblock {Mosip Documentation - Administration}.
\newblock
  \url{https://docs.mosip.io/1.2.0/modules/administration#what-is-deactivation-of-a-resource}.
\newblock (Accessed on February 14, 2024).

\bibitem{mosipdocs2024anonymousprofiling}
{Mosip Documentation}.
\newblock {Mosip Documentation - Anonymous Profiling Support}.
\newblock
  \url{https://docs.mosip.io/1.2.0/id-lifecycle-management/anonymous-profiling-support}.
\newblock (Accessed on February 14, 2024).

\bibitem{mosipdocs2024collabpre}
{Mosip Documentation}.
\newblock {Mosip Documentation - Collab Pre-registration Guide}.
\newblock
  \url{https://docs.mosip.io/1.2.0/collab-getting-started-guide/collab-pre-registration-guide}.
\newblock (Accessed on February 14, 2024).

\bibitem{mosipdocs2024idauthentication}
{Mosip Documentation}.
\newblock {Mosip Documentation - ID Authentication}.
\newblock \url{https://docs.mosip.io/1.2.0/id-authentication}.
\newblock (Accessed on September 22, 2023).

\bibitem{mosipdocs2024overview}
{Mosip Documentation}.
\newblock Mosip documentation - overview.
\newblock \url{https://docs.mosip.io/1.2.0/overview}.
\newblock (Accessed on February 14, 2024).

\bibitem{mosipdocs2024registrationclient}
{Mosip Documentation}.
\newblock Mosip documentation - registration client home page.
\newblock
  \url{https://docs.mosip.io/1.2.0/modules/registration-client/registration-client-home-page#new-registration}.
\newblock (Accessed on September 22, 2023).

\bibitem{mukherjee2021fast}
Anit Mukherjee and Shankar Maruwada.
\newblock Fast-tracking development: A building blocks approach for digital
  public goods.
\newblock {\em Center for Global Development. https://www. cgdev.
  org/sites/default/files/fast-tracking-development-digital-publicgoods. pdf},
  2021.

\bibitem{narayanan2008robust}
Arvind Narayanan and Vitaly Shmatikov.
\newblock Robust de-anonymization of large sparse datasets.
\newblock In {\em 2008 IEEE Symposium on Security and Privacy (sp 2008)}, pages
  111--125. IEEE, 2008.

\bibitem{nist_oscal}
{National Institute of Standards and Technology}.
\newblock Oscal - open security controls assessment language.
\newblock \url{https://pages.nist.gov/OSCAL/}, n.d.
\newblock (Accessed on February 14, 2024).

\bibitem{nicholson2022agenda}
Brian Nicholson, Petter Nielsen, Johan~Ivar S{\ae}b{\o}, and Ana~Paula Tavares.
\newblock Digital public goods for development: A conspectus and research
  agenda.
\newblock In {\em International Conference on Social Implications of Computers
  in Developing Countries}, pages 455--470. Springer, 2022.

\bibitem{nicholson2022digital}
Brian Nicholson, Petter Nielsen, Sundeep Sahay, and Johan~Ivar S{\ae}b{\o}.
\newblock Digital public goods platforms for development: The challenge of
  scaling.
\newblock {\em The Information Society}, 38(5):364--376, 2022.

\bibitem{PIA}
Department of~Homeland~Security.
\newblock Privacy office official guidance for privacy impact assessments.
\newblock
  \url{https://www.dhs.gov/sites/default/files/publications/privacy_pia_guidance_june2010_0.pdf}.
\newblock (Accessed on February 10, 2024).

\bibitem{nist2024privacyframework}
National~Institute of~Standards and Technology.
\newblock Nist privacy framework.
\newblock \url{https://www.nist.gov/privacy-framework/privacy-framework}, Jan
  2024.

\bibitem{onetrust}
OneTrust.
\newblock Pia and dpia automation.
\newblock \url{https://www.onetrust.com/products/pia-and-dpia-automation/}.
\newblock (Accessed on May 23, 2024).

\bibitem{osano}
Osano.
\newblock Data privacy assessment tool.
\newblock \url{https://www.osano.com/products/privacy-assessments}.
\newblock (Accessed on May 23, 2024).

\bibitem{o2020intercoder}
Cliodhna O’Connor and Helene Joffe.
\newblock Intercoder reliability in qualitative research: debates and practical
  guidelines.
\newblock {\em International journal of qualitative methods},
  19:1609406919899220, 2020.

\bibitem{pearson2024breach}
Jordan Pearson.
\newblock The breach of a face recognition firm reveals a hidden danger of
  biometrics, May 2024.

\bibitem{DPI}
United Nations~Development Programme.
\newblock Digital public infrastructure.
\newblock \url{https://www.undp.org/digital/digital-public-infrastructure}.
\newblock (Accessed on May 18, 2024).

\bibitem{privacyref2024framework}
Privacy Ref.
\newblock Choosing a privacy framework.
\newblock \url{https://privacyref.com/blog/choosing-a-privacy-framework/},
  2024.
\newblock (Accessed on February 14, 2024).

\bibitem{reuben2016automated}
Jenni Reuben, Leonardo~A Martucci, and Simone Fischer-H{\"u}bner.
\newblock Automated log audits for privacy compliance validation: a literature
  survey.
\newblock {\em Privacy and Identity Management. Time for a Revolution? 10th
  IFIP WG 9.2, 9.5, 9.6/11.7, 11.4, 11.6/SIG 9.2. 2 International Summer
  School, Edinburgh, UK, August 16-21, 2015, Revised Selected Papers 10}, pages
  312--326, 2016.

\bibitem{rosson2019incentivizing}
John~P Rosson, Mason~J Rice, Juan Lopez~Jr, and Robert~David Fass.
\newblock Incentivizing cyber security investment in the power sector using an
  extended cyber insurance framework.
\newblock {\em Homeland Security Affairs}, 15, 2019.

\bibitem{saebo2021digital}
Johan~Ivar S{\ae}b{\o}, Brian Nicholson, Petter Nielsen, and Sundeep Sahay.
\newblock Digital global public goods.
\newblock {\em arXiv preprint arXiv:2108.09718}, 2021.

\bibitem{oauth}
Swarag Sharma and Jevitha KP.
\newblock Security analysis of oauth 2.0 implementation.
\newblock In {\em 2023 Innovations in Power and Advanced Computing Technologies
  (i-PACT)}, 2023.

\bibitem{sirur2018we}
Sean Sirur, Jason~RC Nurse, and Helena Webb.
\newblock Are we there yet? understanding the challenges faced in complying
  with the general data protection regulation (gdpr).
\newblock In {\em Proceedings of the 2nd International Workshop on Multimedia
  Privacy and Security}, pages 88--95, 2018.

\bibitem{sturmer2023digital}
Matthias St{\"u}rmer, Markus~Andreas Tiede, Jasmin~Myriam Nussbaumer, and
  Flurina W{\"a}spi.
\newblock On digital sustainability and digital public goods.
\newblock 2023.

\bibitem{sweeney2000simple}
Latanya Sweeney.
\newblock Simple demographics often identify people uniquely.
\newblock {\em Health (San Francisco)}, 671(2000):1--34, 2000.

\bibitem{tavares2023digital}
Ana~Paula Tavares, Edgar Whitley, Liv~Marte Nordhaug, Johan S{\ae}b{\o},
  Malavika Raghavan, PK~Senyo, and Silvia Masiero.
\newblock Digital public goods and vulnerable populations.
\newblock 2023.

\bibitem{gdpr-article35}
European Union.
\newblock Article 35 of gdpr - data protection impact assessment.
\newblock \url{https://gdpr.eu/article-35-impact-assessment/}.
\newblock (Accessed on February 10, 2024).

\bibitem{coppa}
{United States Government}.
\newblock Children's online privacy protection rule ("coppa").
\newblock
  \url{https://uscode.house.gov/view.xhtml?req=granuleid%3AUSC-prelim-title15-section6501&edition=prelim}.
\newblock (Accessed on February 13, 2024).

\bibitem{ushahidi}
Ushahidi.
\newblock Ushahidi.
\newblock \url{https://www.ushahidi.com}.

\bibitem{ushahidi_data_obfuscation}
{Ushahidi}.
\newblock Data obfuscation.
\newblock \url{https://www-admin.ushahidi.com/support/data-obfuscation}, n.d.
\newblock (Accessed on February 14, 2024).

\bibitem{ushahidi_surveys}
{Ushahidi}.
\newblock Post types.
\newblock
  \url{https://www-admin.ushahidi.com/support/post-types#what-exactly-is-a-survey},
  n.d.
\newblock (Accessed on February 14, 2024).

\bibitem{ushahidi_deployments}
{Ushahidi}.
\newblock Ushahidi deployments.
\newblock \url{https://www.ushahidi.com/in-action/deployments/}, n.d.
\newblock (Accessed on February 14, 2024).

\bibitem{ushahidi_tech_stack}
{Ushahidi}.
\newblock Ushahidi platform developer documentation: Architecture.
\newblock
  \url{https://docs.ushahidi.com/platform-developer-documentation/tech-stack/architecture},
  n.d.
\newblock (Accessed on February 14, 2024).

\bibitem{ushahidi_datasources}
{Ushahidi}.
\newblock Ushahidi platform user manual: Configuring data sources.
\newblock
  \url{https://docs.ushahidi.com/platform-user-manual/3.-configuring-your-deployment/3.4-data-sources},
  n.d.
\newblock (Accessed on February 14, 2024).

\bibitem{ushahidi_twitter}
{Ushahidi}.
\newblock Ushahidi platform user manual: Configuring twitter data source.
\newblock
  \url{https://docs.ushahidi.com/platform-user-manual/3.-configuring-your-deployment/3.4-data-sources/3.4.7-twitter},
  n.d.
\newblock (Accessed on February 14, 2024).

\bibitem{voigt2017eu}
Paul Voigt and Axel Von~dem Bussche.
\newblock The eu general data protection regulation (gdpr).
\newblock {\em A Practical Guide, 1st Ed., Cham: Springer International
  Publishing}, 10(3152676):10--5555, 2017.

\bibitem{wong2023privacy}
Richmond~Y Wong, Andrew Chong, and R~Cooper Aspegren.
\newblock Privacy legislation as business risks: How gdpr and ccpa are
  represented in technology companies' investment risk disclosures.
\newblock {\em Proceedings of the ACM on Human-Computer Interaction},
  7(CSCW1):1--26, 2023.

\bibitem{wright2016enforcing}
David Wright and Paul De~Hert.
\newblock {\em Enforcing privacy: regulatory, legal and technological
  approaches}, volume~25.
\newblock Springer, 2016.

\bibitem{wuyts2015linddun}
Kim Wuyts and Wouter Joosen.
\newblock Linddun privacy threat modeling: a tutorial.
\newblock {\em CW Reports}, 2015.

\bibitem{zabihzadeh2019cultural}
Abbas Zabihzadeh, Mohammad~Ali Mazaheri, Javad Hatami, Mohammad~Reza Nikfarjam,
  Leili Panaghi, and Telli Davoodi.
\newblock Cultural differences in conceptual representation of “privacy”: A
  comparison between iran and the united states.
\newblock {\em The Journal of social psychology}, 159(4):357--370, 2019.

\bibitem{zimmeck2024generalizable}
Sebastian Zimmeck, Eliza Kuller, Chunyue Ma, Bella Tassone, and Joe Champeau.
\newblock Generalizable active privacy choice: Designing a graphical user
  interface for global privacy control.
\newblock {\em PETS}, 2024.

\end{thebibliography}

\appendix
\label{AppA}

\onecolumn

\section{Codebooks}
\label{app:codebooks}

\begin{table}[htp]
    \centering
    \caption{Codebook for Proposed Protection Mechanisms.}
    \vspace{2pt}
    \begin{tabular}{|p{0.25\textwidth}|p{0.65\textwidth}|} \hline 
         \textbf{Code} & \textbf{Definition} \\ \hline 
         Access Control& The nominee proposes the enforcement of access restrictions to the solution and/or personal data based on predefined rules and policies.\\ \hline 
         Commercial/In-house Tools& The nominee proposes the use of commercial or in-house tools for the protection of personal data.\\ \hline 
         Data Strategies& The nominee proposes data-oriented protection strategies (e.g., minimization, anonymization) as a protective measure for personal data. \\ \hline 
         Notify 3rd Party Data Sharing& The nominee describes the sharing of personal data with third parties for secondary use.\\ \hline 
         Security Upgrades/Patches& The nominee proposes taking responsibility for performing regular security updates and patches to protect personal data.\\ \hline 
         Vulnerability Testing& The nominee proposes taking responsibility for performing regular vulnerability scans to ensure protection of personal data.\\ \hline
         Governance Processes/Audits&The nominee proposes taking responsibility for establishing and/or adhering to a governance process to ensure the protection of personal data.\\\hline
         User Control&The nominee proposes implementing various privacy controls to empower users in expressing their privacy preferences effectively (e.g., user consent, data deletion requests)\\\hline
         Data Storage&The nominee proposes secure data storage solution(s) to ensure protection of personal data.\\\hline
         Strong Passwords&The nominee proposes safeguarding access to personal data by implementing robust password requirements.\\\hline
    \end{tabular}
    \label{tab:ppg_codebook}
\end{table}

\begin{table}[htp]
    \centering
    \caption{Codebook for Provided Supporting Material}
    \vspace{2pt}
    \begin{tabular}{|p{0.25\textwidth}|p{0.65\textwidth}|} \hline 
         \textbf{Code}& \textbf{Definition}\\ \hline 
         No Documentation Submitted& The nominee has not submitted any documentation or references, or has submitted expired links, for review.\\ \hline 
         Security/Privacy Docs& The nominee has shared a link to the solution's security and privacy documentation, which comprises either detailed or high-level information about implementation.\\ \hline 
         Compliance Docs& The nominee has shared a link to the solution's compliance documentation (e.g., GDPR).\\ \hline 
         Cookie Policy& The nominee has shared a link to the solution's cookie practices.\\ \hline 
         Privacy Policy& The nominee has shared a link to the solution's privacy practices. Note: while this is a step towards the right direction, it is still not sufficient.\\ \hline
    \end{tabular}
    \label{tab:psd_codebook}
\end{table}

\begin{table}[htp]
    \centering
    \caption{Codebook for Overall Response Quality}
    \vspace{2pt}
    \label{tab:org_codebook}
    \begin{tabular}{|p{0.25\textwidth}|p{0.65\textwidth}|} \hline 
         \textbf{Code}& \textbf{Definition}\\ \hline 
         Security-related Privacy& Discusses security-related privacy controls, such as encryption and access control, without discussing any data-oriented strategies.\\ \hline 
         Security Only& Focuses only on security measures, without addressing any privacy-related strategies, despite PII being collected, stored, and/or processed\\ \hline 
         Partially Addresses Privacy& Addresses certain aspects of privacy but may not cover all aspects comprehensively.\\ \hline 
         Unclear PII Collection& Nominees lack a clear understanding of what personally identifiable information (PII) entails. Responses are either incorrect or unclear.\\ \hline 
         Clarifies Data Ownership& Emphasis that solution developers (nominee) do not claim ownership of any data collected and/or processed by the solution. Places the burden of privacy on solution implementers, neglecting the fact that privacy-by-design principles should have been incorporated during development.\\ \hline 
         Lack of Specificity& Mentions vague terms without explaining the solution's function/capabilities or proposes privacy-protecting solutions without providing specific implementation details.\\ \hline 
         Downplaying Risks& Clarifies that the nominee do not collect data themselves and explicitly state that they do not own any of the data. This clarification may be intended to downplay potential risks associated with protecting user data.\\ \hline 
         Compliance Implies Protection& Claims compliance with data protection regulations such as GDPR; it does not necessarily indicate that privacy-protecting strategies have been implemented.\\ \hline 
         Inconsistent Answer& Response provided directly contradicts the answers given to other questions.\\ \hline 
         Does Not Answer Question& Incorrect response that does not answer the question.\\ \hline
    \end{tabular}
    \end{table}

\begin{table}[htp]
    \centering
    \caption{Codebook for Privacy Component Analysis}
    \vspace{2pt}
    \begin{tabular}{|p{0.25\textwidth}|p{0.65\textwidth}|}
    \hline
         \textbf{Code}& \textbf{Definition}\\\hline
         Regulatory Efforts& The nominee describes privacy related compliance efforts such as self regulation, enforcement mechanisms, privacy documentation and awareness campaigns.\\\hline
         Notice and Consent& Nominees describes user notice and consent mechanisms.\\\hline
         Data Collection Limitation& Nominee ensures that the system only collects data required for the intended purpose and as for long as necessary.\\\hline
         Data Use Limitation& Nominee ensures that the system only processes data needed to satisfy the intended purpose and describe strategies to do so.\\\hline
         User Choice& Nominee ensures that users provided with appropriate and user-friendly choices in relation to collection, use, transfer and disclosure of their personal information.\\\hline
         Data Accuracy& Nominee ensures that data is accurate and up-to date.\\\hline
         Security Safeguards& Nominee describes security safeguards to protect user data.\\\hline
         Privacy by Design& Nominee addresses Privacy by design (PbD) strategies such as anonymization and privacy preserving defaults.\\\hline
    \end{tabular}
    \label{tab:pca_codebook}
\end{table}

\clearpage

\section{Data Flow Diagrams for Case Studies}

\begin{figure*}[htp]
\begin{minipage}{.5\textwidth}
    \centering
    \includegraphics[width=\textwidth]{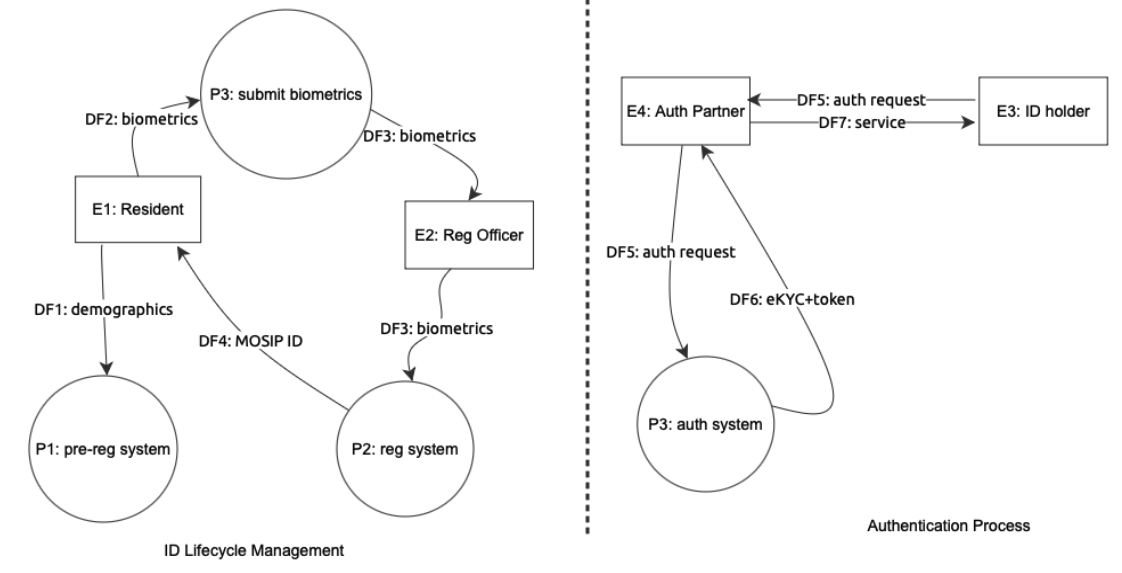}
    \caption{MOSIP's Data Flow Diagram}
    \label{fig:MOSIP_DFD}
\end{minipage}
\begin{minipage}{.5\textwidth}
    \centering
    \includegraphics[width=0.7\textwidth]{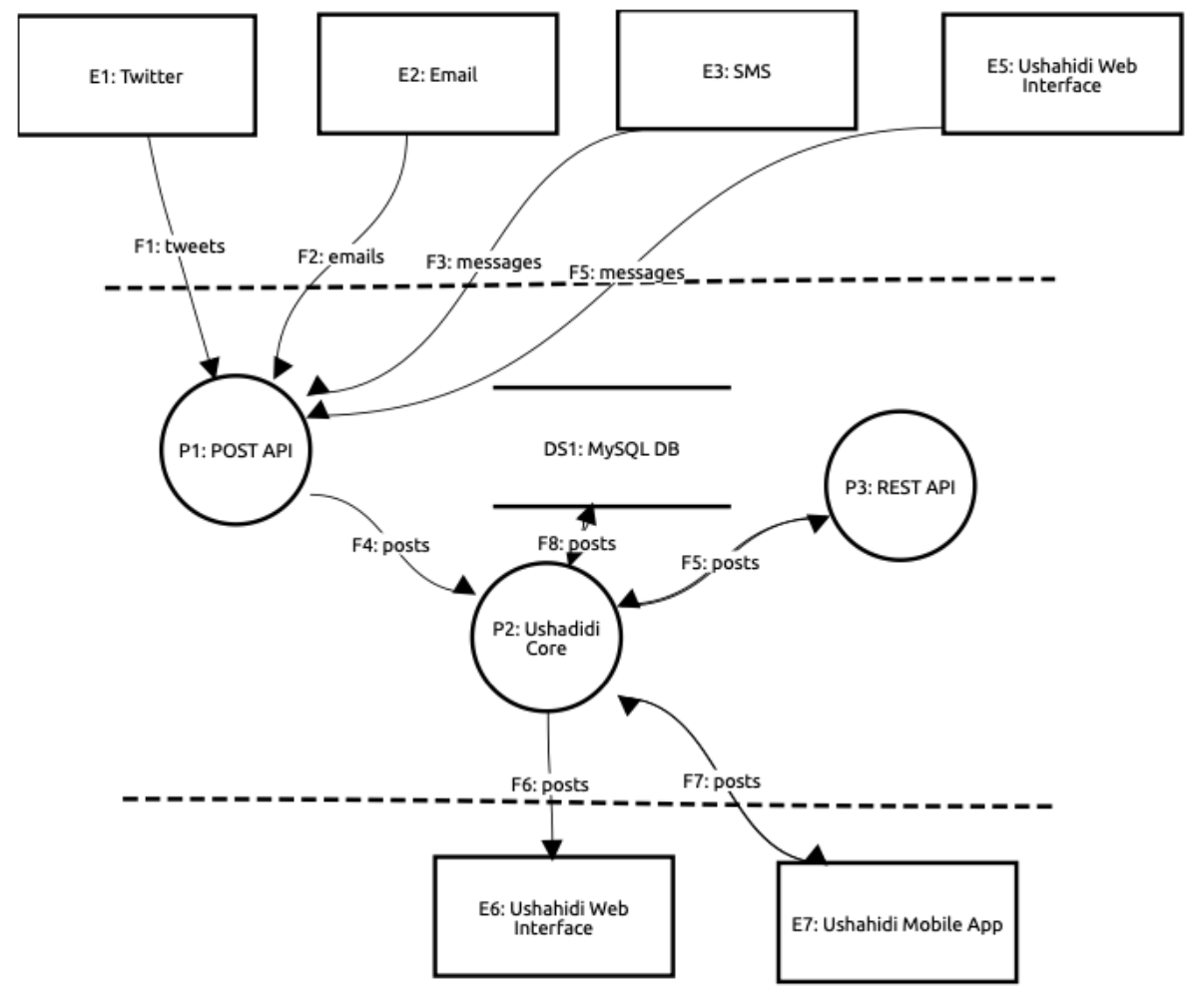}
    \caption{Ushahidi's Data Flow Diagram}
    \label{fig:Ushahidi_DFD}
\end{minipage}
\end{figure*}

\begin{figure*}[htp]
\begin{minipage}{.5\textwidth}
    \centering
    \includegraphics[width=0.8\textwidth]{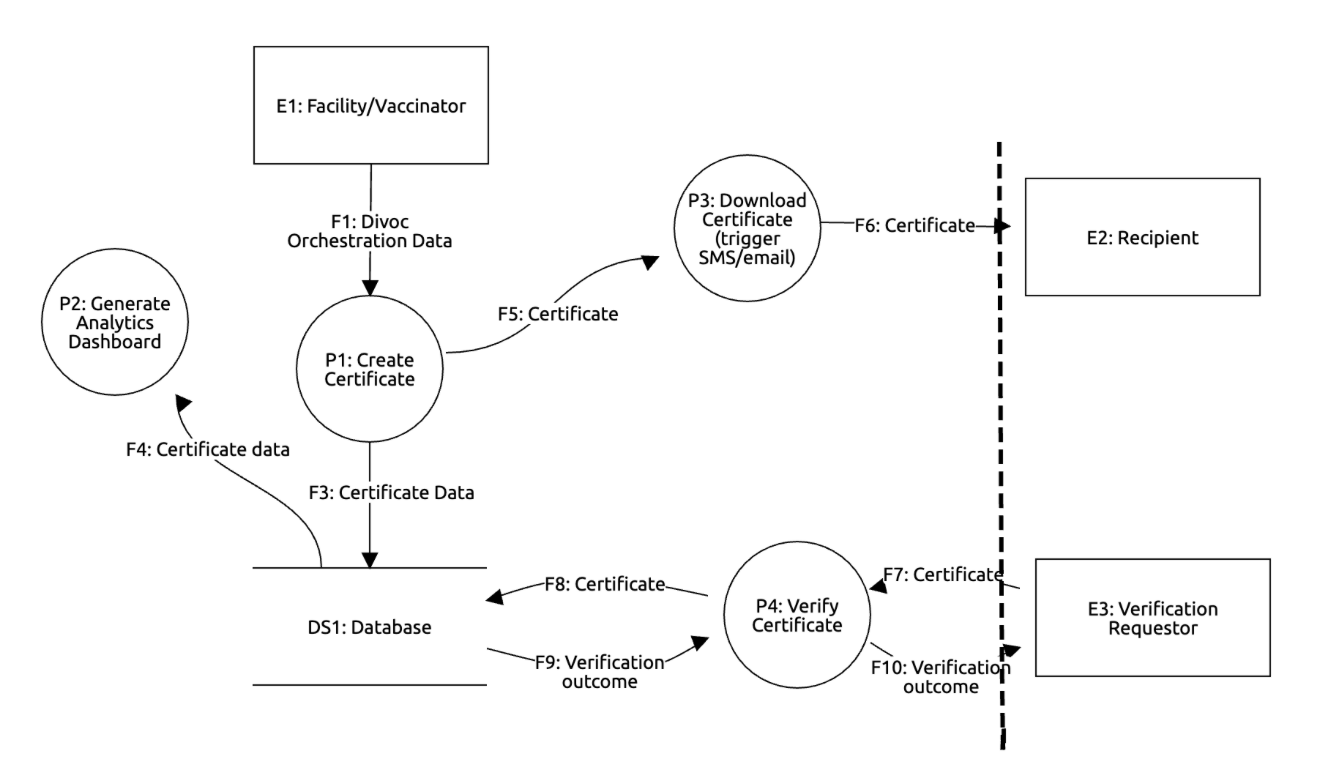}
    \caption{DIVOC's Data Flow Diagram}
    \label{fig:DIVOC_DFD}
\end{minipage}
\end{figure*}

\section{Cost Analysis}
\label{app:cost_analysis}
\update{Choosing between the two tiers in our proposed model will depend on the resources available to the DPGA and the DPG candidate. Self-attestation could impose a burden on DPG candidates, many of which lack privacy and/or compliance teams. Obtaining a certification from approved providers requires less time but possibly more money for DPG candidates, depending on the time cost of completing the assessment and financial compensation of DPG contributors. 
While the proposed model would increase the barrier to DPG certification, we believe basic privacy assessments should be a minimum requirement for organizations handling PII. We summarize the stakeholder cost analysis for our proposed model in Table \ref{tab:stakeholder_analysis}.}

\begin{table}[h]
\centering
\caption{\update{Stakeholder cost analysis of the online cost (i.e., during DPG certification) of the two privacy certification tiers of our proposed model. Arrows indicate the change in resources compared to what is currently needed.}}
\label{tab:stakeholder_analysis}
\begin{tabular}{|p{0.25\linewidth}|p{0.2\linewidth}|p{0.07\linewidth}|p{0.1\linewidth}|p{0.11\linewidth}|} 
\hline
\textbf{Proposed Strategy} & \textbf{Stakeholder} & \textbf{Time} & \textbf{Money} & \textbf{Overall Effort}  \\ 
\hline
\multirow{3}{*}{\shortstack[l]{\\Option 1: \\ PIA by Certified Provider}} & DPGA & $\downarrow$ & $\downarrow$ & $\downarrow$ \\ 
\cline{2-5}
 & Approved Provider & $\uparrow$ & $\downarrow$ & $\uparrow$ \\ 
\cline{2-5}
 & DPG Cand. & $\downarrow$ & $\uparrow$ & $\uparrow$ \\ 
\hline
\multirow{2}{*}{\shortstack[l]{\\ Option 2: \\ Self Attestation}} & DPGA & $\downarrow$ & $\downarrow$ & $\downarrow$  \\ 
\cline{2-5}
 & DPG Cand. & $\uparrow$ & $\uparrow$ & $\uparrow$  \\
\hline
\end{tabular}
\end{table}

\end{document}